\documentclass{iopconfser}
\usepackage[numbers,sort&compress]{natbib}
\usepackage[colorlinks, linkcolor=blue, citecolor = blue, urlcolor = blue]{hyperref}
\usepackage{amsmath}
\usepackage{amssymb}
\usepackage[title, toc]{appendix}
\usepackage{graphicx}
\usepackage{dcolumn}
\usepackage{bm}
\usepackage{caption}
\usepackage{subcaption}
\usepackage{pgfplots}
\usepackage{amsthm}

\newcommand{\kij}{k_{ij}}
\newcommand{\kji}{k_{ji}}
\newcommand{\GM}{G^{\Delta t}} 
\newcommand{\GMT}{\re{G^{\Delta t}}}
\renewcommand{\Re}{\text{Re}\,}
\renewcommand{\Im}{\text{Im}\,}

\newcommand{\esc}[1]{r_{#1}}                     
\newcommand{\lbesc}[1]{r_{#1}^{\text{lb}}}       
\newcommand{\ubesc}[1]{r_{#1}^{\text{ub}}}       

\newcommand{\escmax}{r_{\text{max}}}             
 
\newcommand{\ubescmax}{r^{\text{ub}}_{\text{max}}}

\newcommand{\G}[2]{G^{\Delta t}_{#1#2}}   

\newcommand{\az}{N}

\newcommand{\sigmacgr}{\hat{\sigma}}

\newcommand{\re}[1]{\textcolor{black}{#1}}

\newtheorem*{Gerschgorins_Theorem}{Gerschgorin Circle Theorem \cite{Gerschgorin}}

\begin{document}

\title{Stroboscopic measurements in Markov networks: Exact generator reconstruction vs. thermodynamic inference}

\author{Malena T. Bauer$^{1}$, Udo Seifert$^{1}$ and Jann van der Meer$^{2}$}

\affil{$^1$II. Institut für Theoretische Physik, Universität Stuttgart, 70550 Stuttgart, Germany}
\affil{$^2$Department of Physics \#1, Graduate School of Science, Kyoto University, Kyoto 606-8502, Japan}

\email{vandermeer.jann.5t@kyoto-u.ac.jp}

\begin{abstract}
A major goal of stochastic thermodynamics is to estimate the inevitable dissipation that accompanies particular observable phenomena in an otherwise not fully accessible system. Quantitative results are often formulated as lower bounds on the total entropy production, which capture the part of the total dissipation that can be determined based on the available data alone. In this work, we discuss the case of a continuous-time dynamics on a Markov network that is observed stroboscopically, i.e., at discrete points in time in regular intervals. We compare the standard approach of deriving a lower bound on the entropy production rate in the steady state to the less common method of reconstructing the generator from the observed propagators by taking the matrix logarithm. Provided that the timescale of the stroboscopic measurements is smaller than a critical value that can be determined from the available data, this latter method is able to recover all thermodynamic quantities like entropy production or cycle affinities and is therefore superior to the usual approach of deriving lower bounds. Beyond the critical value, we still obtain tight upper and lower bounds on these quantities that improve on extant methods. We conclude the comparison with numerical illustrations and a discussion of the requirements and limitations of both methods.
\end{abstract}

\section{Introduction}

The framework of stochastic thermodynamics provides a theoretical foundation for thermodynamic quantities on the nanoscale and the physical laws relating them \cite{seif12, vdb15, peli21, shir23}. Building upon this groundwork, the emerging subfield of stochastic thermodynamics that has been dubbed ``thermodynamic inference'' investigates how such fundamental principles can be utilized to gain insights into experimentally accessible situations \cite{seif19}. A common formulation of the problem of thermodynamic inference is to ask what the principles of stochastic thermodynamics can reveal about a physical system and, in particular, about characteristic thermodynamic quantities like its entropy production in the unfavorable but common situation that not all relevant parts of the system can be observed directly.

A prominent early result that strongly influenced the field of thermodynamic inference is the thermodynamic uncertainty relation (TUR) \cite{bara15, ging16, horo20}, which provides a lower bound on the entropy production of a discrete or continuous Markovian dynamics in terms of the precision of an externally observable current. Gradually becoming an archetype for a considerable number of subsequent results in thermodynamic inference, such lower bounds on the total entropy production in terms of an operationally accessible quantity establish that the observation of particular phenomena necessarily implies a certain amount of dissipation. A fairly general way to obtain inequalities of this form is to interpret the entropy production as a Kullback-Leibler divergence \cite{kawa07, gome08a, gome08b, rold10} and utilize information-theoretic inequalities \cite{cove06}. This technique has proven a flexible template to establish estimators for entropy production in partially observed discrete Markov networks \cite{espo12, seif19}, which often also satisfy a fluctuation relation on the coarse-grained level \cite{shir14, bisk17, pole17a, degu23}, or quantify the irreversibility that is apparent from the coarse-grained semi-Markov model \cite{mart19, vdm22, haru22, vdm23, blom24, degu24, kapu24}. In contrast to such methods that identify an entire coarse-grained dynamics, lower bounds on entropy production that focus on signatures of nonequilibrium in particular observable quantities have been formulated based on, for example, waiting times within lumped states \cite{skin21a} or between time-symmetric counting events \cite{piet24}, transition statistics \cite{skin21}, antisymmetric cross-correlations \cite{ohga23, shir23b, vu23a}, correlation times \cite{dech23},  or spectral methods \cite{kolc23}. 

However and despite the diversity of the previously discussed methods, there is no fundamental reason to limit results of thermodynamic inference to lower bounds on entropy production. Assuming a particular model class for the physical system and its dynamics can already impose strong restrictions on the hidden parts of the system, which are perhaps not taken into account in an optimal way if one considers only a lower bound on a single averaged quantity like entropy production. In this work, we face such a situation by assuming that an underlying dynamics that evolves on a Markov network is monitored only at discrete points in time. More precisely, we assume that the state of the system is measured in regular time intervals of length $\Delta t$, so that coarse graining yields a Markov process in discrete time. 

Given that such a temporal coarse graining is unavoidable when dealing with data from experimental signals, as discussed, e.g., in Refs. \cite{skin21, luce23}, studies regarding the estimation of entropy production in discrete-time Markov processes are surprisingly scarce compared to the case of continuous time. Perhaps, this is due to the implicit assumption that for sufficiently small $\Delta t$ the more tractable continuous-time model captures all essential features. One of the rare findings that applies to discrete-time processes in particular is the curious result that the usual formulation of the TUR can be violated \cite{shir17a} and has to be replaced by a weaker but more general bound \cite{proe17}. The relation between TURs for continuous-time and associated discrete-time Markov processes has been investigated in more detail in Ref. \cite{chiu18}. On a more formal level, results that are established for semi-Markov processes also include the case of discrete time, which allows us not only to understand the discrete-time TUR as a more general result valid for stationary semi-Markov processes \cite{erte22}, but also to establish relations like the fluctuation theorem \cite{andr08b}.

Taking into account that the coarse-grained dynamics emerges from an underlying continuous-time Markov process, we can apply further results of thermodynamic inference to, e.g., estimate entropy production \cite{rold10, vdm23, cisn23} or cycle affinities \cite{lian23, degu23} in the stationary state. However, merely specializing these general results to discrete time seems suboptimal, at least for sufficiently small (but finite) $\Delta t$, because if the propagator is close enough to the identity matrix it contains all the necessary information to recover the associated generator (e.g. by calculating the matrix logarithm through a power series). Further investigations of this systematic way to exactly recover the generator from stroboscopically observed data will also shed light on the conditions under which this method outperforms the more usual approaches to thermodynamic inference.

On a different note, we point out that interest in reconstructing the generator of a discrete Markovian dynamics is not limited to stochastic thermodynamics. The question of whether the underlying generator of a continuous-time Markov process can be unambiguously recovered from the propagator for finite $\Delta t$ has been discussed in various other contexts before and dates back almost as long as the theory of Markov chains itself \cite{elfv37}. This question and closely related problems have been studied not only in some older mathematical literature, e.g. Refs. \cite{king62, cuth72, cuth73, joha74}, but also in the context of financial mathematics \cite{isra01}, social studies \cite{cole73, sing73, sing76}, statistics and computational methods \cite{blad05, metz07, metz07a}, and biochemistry \cite{verb13, Equilibrium_embedding_problem}, in particular molecular dynamics \cite{spec19}. Some of the results in this paper have already been derived in the respective context in one of these works. In this paper, we compile these results and complement them with additional, original findings through the lens of thermodynamic inference. For clarity, we adopt a self-contained approach in this work, which will not only allow insight into the results applied here and how to strengthen them but also into how these findings relate to more common methods of thermodynamic inference.

The paper is structured as follows. We define the set-up and central quantities of this work in Section \ref{sec:2}. The subsequent Section \ref{sec:3} introduces the matrix logarithm to obtain the generator of the underlying process. The section includes proofs of the main statements to allow for a self-contained presentation. We compare this method to more conventional results of thermodynamic inference like estimators for entropy production and affinities in Section \ref{sec:4}. The concluding Section \ref{sec:5} discusses our findings in the context of thermodynamic inference and outlines potential future work.

\section{Set-up} \label{sec:2}
We consider a discrete Markovian system with $\az$ states and time-independent rates $\kij$ for a transition from state $i$ to state $j$. We denote the escape rate at which an occupied state $i$ is exited by
\begin{equation}
    \esc{i}\equiv \sum_{j\neq i}\kij
\end{equation}
and refer to the maximal escape rate as
\begin{equation}
\escmax\equiv \max_i \esc{i}.
\end{equation}
By defining the generator matrix $L_0$ with matrix elements $\left( L_0 \right)_{ij}$ as
\begin{equation}
    \left( L_0 \right)_{ij} =\begin{cases}
        \kji&\text{for }i\neq j\\
        -\esc{i}&\text{for }i= j
    \end{cases}
\end{equation}
the time-evolution of such a system is governed by the master equation
\begin{equation}
\label{eq:master_eq}
    \partial_t p=L_0 p
\end{equation}
for the $N$-dimensional probability distribution $p(t)$. We assume that the underlying network is connected and that the system is in its unique stationary state $p^s$, i.e., $p(t)=p^{\text{s}}$ with $L_0 p^{\text{s}}=0$. The mean entropy production rate is then given by \cite{seif12}
\begin{equation}
\label{eq:mean_entropy}
    \sigma=\sum_{i\neq j}p_i^{\text{s}}\kij\ln\left(\frac{\kij}{\kji}\right).
\end{equation}
The affinity of a cycle $\mathcal{C}$ in the system is another quantity of thermodynamic importance and defined as
\begin{equation}
\label{eq:affinity}
    \mathcal{A}_{\mathcal{C}}=\ln\left(\prod_{(ij)\in\mathcal{C}}\frac{\kij}{\kji}\right)
,\end{equation}
where the product runs over all transitions along a particular direction in a cycle.

We assume that the state of the system system is measured stroboscopically with a fixed interval of length $\Delta t$ between two consecutive observations. In the following, we will assume the idealized scenario of in principle infinite data, which allows us to infer the probability of finding the system in state $\re{i}$ at time $t=\Delta t$ given that it was in state $\re{j}$ at time $t=0$ for all states $i,j$. These conditioned probabilities define the propagators
\begin{equation}
    \re{\G{i}{j}\equiv p(i,\Delta t|j,0)}.
\end{equation}
The matrix with entries $\G{i}{j}$ will be denoted as $\GM$ in the following. 
\re{This matrix is directly linked to the dynamics via}
\begin{equation}
    \GMT=e^{\Delta t L_0},
\end{equation}
because every column of $\GMT$ solves the master equation \eqref{eq:master_eq}.

Given sufficient data from stroboscopic measurements, we are able to obtain $\GM$ and from this matrix also the stationary probability distribution $p^{\text{s}}$. Thus, we can consider the inference problem of reconstructing the unknown generator matrix $L_0$ from these two quantities. As an immediate consequence of their definition, generator matrices of physical Markovian systems are real, have non-negative entries except on the diagonal and are column-stochastic, i.e.,
\begin{equation}
\label{eq:def_column_stochastic}
    \sum_i L_{ij}=0
\end{equation}
for all $j$. In this thermodynamic setting of inference, we additionally require that $L_{ij}>0$ implies $L_{ji}>0$ for all states $i\neq j$, which is needed to avoid infinite entropy production (cf. Eq. \eqref{eq:mean_entropy}). A matrix satisfying these properties will be referred to as a ``permissible generator matrix" in the following. \re{We note that the condition $L_{ij}>0 \Rightarrow L_{ji}>0$, which is sometimes referred to as microscopic reversibility, is necessary to relate transition rates and thermodynamic quantities, see e.g., Ref. \cite{seif12}.} 

\section{Inference via generator reconstruction} \label{sec:3}
In this section, we investigate whether it is possible to determine the unknown generator matrix $L_0$ from the given matrix $\GMT=\exp(\Delta t L_0)$ for a particular, fixed $\Delta t$. We adopt two different approaches. 

\re{We first consider conditions on $\Delta t$ that allow us to reconstruct the generator uniquely from the given data. Our reasoning in the according Sections \ref{sec:small_dt} and \ref{sec:upper_bound_r_max} is roughly equivalent to the arguments provided in Ref. \cite{cuth73}, but we aim at a presentation that is appropriate for the purposes of inference.}

In a complementary second approach, we consider arbitrary $\Delta t$. \re{We first discuss an important class of propagators in which unique reconstruction of the generator is always possible. This finding and related results are discussed in more detail in Ref. \cite{sing76}}. In the more general case of finitely many possible generators, we show how to obtain tight upper and lower bounds on thermodynamic quantities\re{, which constitutes an original contribution and is one of our main results.}

\subsection{Uniqueness of the generator for sufficiently small spacings between observations}
\label{sec:small_dt}

We first derive a result that allows us to reconstruct the unique generator, provided that $\Delta t$ is smaller than a particular value. Although this result may not seem immediately useful, it will serve as the basis for our subsequent findings. 

\re{A central element of the proof is the Gerschgorin circle theorem \cite{Gerschgorin}, which constrains the spectrum of a matrix.}

\re{\begin{Gerschgorins_Theorem}
Let $A\in\mathbb{C}^{N\times N}$ be an arbitrary matrix with entries $a_{ij}$ and eigenvalues $z_k$. We define the set
\begin{equation}
    \mathcal{G}\equiv \bigcup_i \{x\in\mathbb{C}:|x-a_{ii}|\leq R_i\}
\end{equation}
with
\begin{equation}
    R_i\equiv \sum_{j\neq i}|a_{ij}|.
\end{equation}
Then all eigenvalues lie in this set, i.e., $z_k\in \mathcal{G}$ for all $k$. We will refer to a set $\{x\in\mathbb{C}:|x-a_{ii}|\leq R_i\}$ as a Gerschgorin circle.
\end{Gerschgorins_Theorem}}

\re{The following statement will apply the previous theorem to permissible generator matrices. In this case, all eigenvalues lie in a disc whose center lies on the real axis. We combine this fact with known properties about the matrix logarithm \cite{Buch_Matrix_log} to formulate a uniqueness statement about a generator matrix satisfying particular properties.}

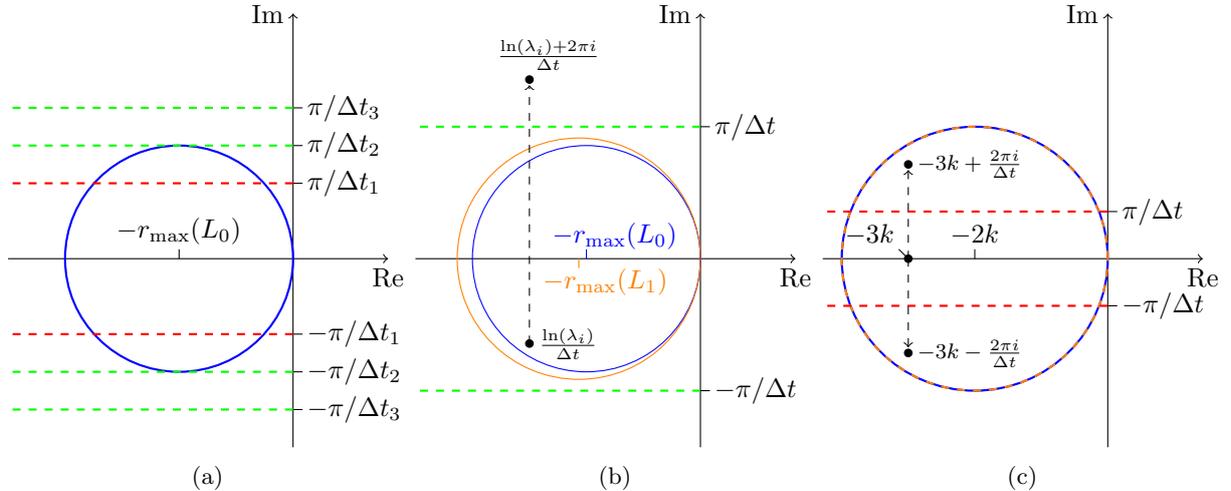
\begin{figure}[t]
\begin{minipage}{0.33\textwidth}
    \centering
    \begin{tikzpicture}[scale=2.5]
    \draw[black,->](-1.5,0)--(0.5,0);
    \draw[black,->](0,-1)--(0,1.3);
    \draw(0.5,0) node[anchor=north]{Re};
    \draw(0,1.3) node[anchor=east]{Im};
    \draw[blue,thick](-0.6,0)circle(0.6);
        \draw[black](0.05,0.4)--node[anchor=west]{$\pi/\Delta t_1$}(0,0.4);
    \draw[black](0.05,-0.4)--node[anchor=west]{$-\pi/\Delta t_1$}(0,-0.4);
    \draw[red,dashed,thick](0,0.4)--(-1.5,0.4);
    \draw[red,dashed,thick](0,-0.4)--(-1.5,-0.4);
    \draw[black](0.05,0.6)--node[anchor=west]{$\pi/\Delta t_2$}(0,0.6);
    \draw[black](0.05,-0.6)--node[anchor=west]{$-\pi/\Delta t_2$}(0,-0.6);
    \draw[green,dashed,thick](0,0.6)--(-1.5,0.6);
    \draw[green,dashed,thick](0,-0.6)--(-1.5,-0.6);
        \draw[black](0.05,0.8)--node[anchor=west]{$\pi/\Delta t_3$}(0,0.8);
    \draw[black](0.05,-0.8)--node[anchor=west]{$-\pi/\Delta t_3$}(0,-0.8);
    \draw[green,dashed,thick](0,0.8)--(-1.5,0.8);
    \draw[green,dashed,thick](0,-0.8)--(-1.5,-0.8);
    \draw[black](-0.6,0)--node[anchor=south]{$-\escmax(L_0)$}(-0.6,0.05);
\end{tikzpicture}
\subcaption{}
    \label{fig:visualization_proof_different_dt}
    \end{minipage}
    \begin{minipage}{0.33\textwidth}
    \centering
\begin{tikzpicture}[scale=2.5]
    \draw[black,->](-1.5,0)--(0.5,0);
    \draw[black,->](0,-1)--(0,1.3);
    \draw(0.5,0) node[anchor=north]{Re};
    \draw(0,1.3) node[anchor=east]{Im};
    
    \draw[blue](-0.6,0)circle(0.6);
    
    \draw[blue](-0.6,0)--(-0.6,0.05);
    \draw(-0.45,0)[blue] node[anchor=south]{$-\escmax(L_0)$};
    
    \draw[orange](-0.64,0)circle(0.64);
    \draw[orange](-0.64,0)--(-0.64,-0.05);
    \draw(-0.5,0)[orange] node[anchor=north]{$-\escmax(L_1)$};
    
        \draw[black](0.05,0.7)--node[anchor=west]{$\pi/\Delta t$}(0,0.7);
    \draw[black](0.05,-0.7)--node[anchor=west]{$-\pi/\Delta t$}(0,-0.7);
    \draw[green,dashed,thick](0,0.7)--(-1.5,0.7);
    \draw[green,dashed,thick](0,-0.7)--(-1.5,-0.7);
    \filldraw[black](-0.9,-0.45)circle (0.02);
    \draw[black](-0.9,-0.45) node[anchor=west]{\footnotesize$\frac{\text{ln}(\lambda_i)}{\Delta t}$};
    \filldraw[black](-0.9,0.95)circle (0.02);
    \draw[black](-0.8,0.95) node[anchor=south]{\footnotesize$\frac{\text{ln}(\lambda_i)+2\pi i}{\Delta t}$};
    \draw[black,dashed,->](-0.9,-0.45)--(-0.9,0.92);
\end{tikzpicture}
    \subcaption{}
    \label{fig:proof_inequ_satisfied}
\end{minipage}
\begin{minipage}{0.33\textwidth}
    \centering
    \begin{tikzpicture}[scale=2.5]
    \draw[black,->](-1.5,0)--(0.5,0);
    \draw[black,->](0,-1)--(0,1.3);
    \draw(0.5,0) node[anchor=north]{Re};
    \draw(0,1.3) node[anchor=east]{Im};
    
    \draw[blue,thick](-0.7,0)circle(0.7);
    \draw[orange,dashed,thick](-0.7,0)circle(0.7);
    
        \draw[black](0.05,0.25)--node[anchor=west]{$\pi/\Delta t$}(0,0.25);
    \draw[black](0.05,-0.25)--node[anchor=west]{$-\pi/\Delta t$}(0,-0.25);
    \draw[red,dashed,thick](0,0.25)--(-1.5,0.25);
    \draw[red,dashed,thick](0,-0.25)--(-1.5,-0.25);

    \draw[black](-0.7,0)--node[anchor=south]{$-2k$}(-0.7,0.05);
    \filldraw[black](-1.05,0)circle (0.02);
    \draw[black](-1.05,0)--node[anchor=south east]{$-3k$}(-1.1,0.05);

    \filldraw[black](-1.05,0.5)circle (0.02);
    \draw[black](-1.05,0.5) node[anchor=west]{\footnotesize$-3k+\frac{2\pi i}{\Delta t}$};
    \draw[black,dashed,->](-1.05,0)--(-1.05,0.47);
    
    \filldraw[black](-1.05,-0.5)circle (0.02);
    \draw[black](-1.05,-0.5) node[anchor=west]{\footnotesize$-3k-\frac{2\pi i}{\Delta t}$};
    \draw[black,dashed,->](-1.05,0)--(-1.05,-0.47);
\end{tikzpicture}
    \subcaption{}
    \label{fig:example_two_different_generators}
\end{minipage}
    \caption{Visualization of the proof from Section \ref{sec:small_dt}. 
    (a) Gerschgorin circle around $-\escmax(L_0)$ of a permissible generator matrix $L_0$ (cf. Section \ref{sec:small_dt}). The horizontal lines show three different possible values of $\pi/\Delta t$, with $\Delta t$ as the time interval between two consecutive observations. For the largest value $\Delta t_1$ the inequality $\escmax(L_0)<\pi/\Delta t_1$, i.e. the condition \eqref{condition:r_smaller_pi_over_dt}, is violated. In this case, it is possible that another permissible generator matrix $L_1\neq L_0$ with $\exp(\Delta t_1 L_1)=\exp(\Delta t_1 L_0)$ exists, and in this case we cannot draw any conclusions about the maximal escape rate $\escmax(L_1)$. For $\Delta t_2$ and $\Delta t_3$, such a scenario is not possible: Any permissible generator matrix $L_1\neq L_0$ with $\exp(\Delta t_1 L_1)=\exp(\Delta t_1 L_0)$ must violate condition \eqref{condition:r_smaller_pi_over_dt}. 
    (b) The favorable scenario that $\escmax(L_0)\leq \pi/\Delta t$ is satisfied, which corresponds to the two pairs of dashed green lines in Figure a). In this case, adding $2\pi i/\Delta t$ to an eigenvalue of $L_0$ moves it out of the Gerschgorin circle (cf. Eq. \eqref{inequ:IM_mu_geq_pi_over_dt}). 
    A hypothetical matrix $L_1$ that has this modified eigenvalue cannot be a permissible generator matrix that satisfies $\escmax(L_1) \leq \pi/\Delta t$, which is a consequence of the Gerschgorin circle theorem. 
    (c) The unfavorable scenario that $\escmax(L_0) > \pi/\Delta t$, which corresponds to the pair of dashed red lines in Figure a). In Sec. \ref{sec:real_eigenvalues} we give an explicit example where $\escmax(L_0)=\escmax(L_1)=2k$, i.e. the largest Gerschgorin circles of the two matrices are identical. For $\Delta t>2\pi/(\sqrt{3}k)$, $L_1$ is a permissible generator matrix. As illustrated in the figure, this is not in contradiction to the Gerschgorin circle theorem, because the shifted eigenvalues do not leave the Gerschgorin circle around $-\escmax(L_1)$.}
    \label{fig:proof_different_scenarios_gerschgorin_circles}
\end{figure}

\paragraph{Statement.}
Let $L_0$ be a permissible generator matrix with the properties
\begin{subequations}
    \begin{equation}
    \exp(\Delta t L_0)=\GMT\label{eq:exp_dt_L_0_GMT}
    \end{equation}
and
    \begin{equation}
    \escmax(L_0)<\frac{\pi}{\Delta t}.\label{condition:r_smaller_pi_over_dt}
    \end{equation}
\end{subequations}
Then $L_0$ is the only such matrix and can be determined from $\GM$ constructively. The final expression for $L_0$ given in Eq. \eqref{eq:L_0_Z_lnD_Z_invers} will be derived in the following proof. A visualization of the proof is shown in Figure \ref{fig:proof_different_scenarios_gerschgorin_circles}.

Put colloquially, the result states that we can successfully reconstruct the generator of the system from the matrix of propagators $\GM$, provided that we knew somehow that the spacing $\Delta t$ between the observations is smaller than $\pi/\escmax$. Since there is no reason to assume that $\escmax$ should be known to an external observer, in Section \ref{sec:upper_bound_r_max} we complement this result with \re{a variant proved by Cuthbert \cite{cuth73}, which} requires a more restrictive condition than \eqref{condition:r_smaller_pi_over_dt}, but does not require knowing the maximal escape rate.

\paragraph{Proof and remarks.}
We assume that the matrix $\GMT$ is diagonalizable in the form
\begin{equation}
    \GMT= ZDZ^{-1}
\end{equation}
with $D=\text{diag}(\lambda_1,...,\lambda_{\az})$ and an invertible matrix $Z$. The nondiagonalizable case is discussed briefly at the end of this proof.

According to Theorem 1.27 in Ref. \cite{Buch_Matrix_log}, all solutions of Eq. \eqref{eq:exp_dt_L_0_GMT}, i.e., all matrices $L_J$ with $\exp(\Delta t L_J)=\GMT$, have the form 
\begin{equation}
\label{eq:matrix_log_def}
    L_J=ZU\left(\frac{1}{\Delta t} \ln(D) +\frac{2\pi i}{\Delta t}\text{diag}(j_1,...,j_{\az}) \right)U^{-1}Z^{-1},
\end{equation}
with arbitrary $J\equiv(j_1,...,j_{\az})\in \mathbb{Z}^{\az}$, $U$ an arbitrary invertible matrix that commutes with $D$, and $\ln(D)\equiv \text{diag}(\ln(\lambda_1),...,\ln(\lambda_{\az}))$, where $\ln$ denotes the principal branch of the logarithm with $\text{Im}(\ln(x))\in (-\pi,\pi]$.

Any matrix $L_J$ of the form \eqref{eq:matrix_log_def} satisfies condition \eqref{eq:exp_dt_L_0_GMT} by construction. We now assume that $L_J$ also satisfies the second condition \eqref{condition:r_smaller_pi_over_dt} and is a permissible generator matrix. It remains to show that only one such matrix $L_J$ exists. 

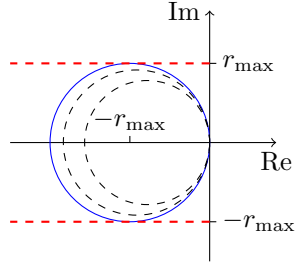
\begin{figure}[t]
\centering
    \begin{tikzpicture}[scale=1.75]
    \draw[black,->](-1.5,0)--(0.5,0);
    \draw[black,->](0,-0.9)--(0,1.0);
    \draw(0.5,0) node[anchor=north]{Re};
    \draw(0,1.0) node[anchor=east]{Im};
    \draw[blue](-0.6,0)circle(0.6);
    \draw[black,dashed](-0.55,0)circle(0.55);
    \draw[black,dashed](-0.47,0)circle(0.47);
    \draw[black](0.05,0.6)--node[anchor=west]{$\escmax$}(0,0.6);
    \draw[black](0.05,-0.6)--node[anchor=west]{$-\escmax$}(0,-0.6);
    \draw[black](-0.6,0)--node[anchor=south]{$-\escmax$}(-0.6,0.05);
    \draw[red,dashed,thick](0,0.6)--(-1.5,0.6);
    \draw[red,dashed,thick](0,-0.6)--(-1.5,-0.6);
\end{tikzpicture}
    \caption{Three Gerschgorin circles of a permissible generator matrix. Since the insides of the two dashed circles are subsets of the inside of the blue circle with radius $\escmax$, all eigenvalues lie in this circular area. In particular, they have to lie between the two red lines, i.e, the absolute value of the imaginary part of all eigenvalues has to be smaller than or equal to the radius $\escmax$.}
    \label{fig:gerschgorin_circles}
\end{figure}

First, we show that $J=(0,...,0)$ has to hold under the given assumptions. According to Eq. \eqref{eq:matrix_log_def} the eigenvalues of $L_J$ are given by 
\begin{equation}
\label{eq:sec2:mu_i}
\mu_i^{(J)}\equiv \left[ \ln(\lambda_i)+2\pi i j_i)  \right]/\Delta t,
\end{equation}
i.e., the imaginary parts of these eigenvalues depend on $J$. Under the given conditions, these imaginary parts can be constrained by the Gerschgorin circle theorem.
\re{We apply this theorem to $L_J^{\text{T}}$, which has the same eigenvalues as $L_J$. Since $L_J$ is a permissible generator matrix, we can write the quantities that are relevant for the theorem in terms of the escape rates as $R_i=-\left(L_J^{\text{T}}\right)_{ii}=\esc{i}(L_J)$.}
Since all Gerschgorin circles are nested inside the largest one, all eigenvalues of $L_J$ lie within a circle with radius $\escmax(L_J)$ around the center $-\escmax(L_J)$, as illustrated in Figure \ref{fig:gerschgorin_circles}.
As a consequence, we have
\begin{equation}
\label{ineq:Im_mu_leq_r_max}
    |\text{Im}(\mu_i^{(J)})|\leq \escmax(L_J)< \pi/\Delta t
\end{equation}
for all $i$, where the last inequality follows from condition \eqref{condition:r_smaller_pi_over_dt}. If $J\neq (0,..,0)$, there exists an $i$ with $|j_i|\geq 1$ and therefore
\begin{eqnarray}
\label{inequ:IM_mu_geq_pi_over_dt}
    |\text{Im}(\mu_i^{(J)})|&&=\frac{1}{\Delta t} |\text{Im}(\ln(\lambda_i)+2\pi i j_i)|\nonumber\\
    &&\geq \frac{1}{\Delta t} \big| |2\pi  j_i|-|\text{Im}(\ln(\lambda_i))| \big|\nonumber\\
   && \geq \frac{\pi}{\Delta t},
\end{eqnarray}
since $|\text{Im}(\ln(\lambda_i))|\leq \pi$. This contradiction to inequality \eqref{ineq:Im_mu_leq_r_max} suffices to conclude $J=(0,...,0)$ if the permissible generator $L_J$ satisfies condition \eqref{condition:r_smaller_pi_over_dt}.

So far, we know that all permissible generator matrices $L$ that satisfy the conditions of the theorem have the form
\begin{equation}
\label{eq:result_L_with_U}
    L= \frac{1}{\Delta t} ZU\ln(D)U^{-1}Z^{-1}
\end{equation}
with an arbitrary invertible matrix $U$ that commutes with $D$. However, since $U$ commutes with $D$ it also commutes with $\ln(D)$ and therefore does not affect the matrix $L$. Thus, the matrix
\begin{equation}
\label{eq:result_L_without_U}
    L= \frac{1}{\Delta t} Z\ln(D)Z^{-1}
\end{equation}
is the only permissible generator matrix that satisfies the conditions \eqref{eq:exp_dt_L_0_GMT} and \eqref{condition:r_smaller_pi_over_dt}. Since the generator matrix $L_0$ of the observed system is a permissible generator matrix that satisfies these conditions by assumption, we obtain
\begin{equation}
\label{eq:L_0_Z_lnD_Z_invers}
    L_0=L= \frac{1}{\Delta t} Z\ln(D)Z^{-1}
\end{equation}
and the statement is proven.

In the case of nondiagonalizable $\GMT$ the proof proceeds similarly, with two modifications. First, the diagonal matrix $D$ has to be replaced with the Jordan canonical form of $\GMT$. We also have to use the general form of the matrix logarithm described in Theorem 1.27 of Ref. \cite{Buch_Matrix_log} instead of Eq. \eqref{eq:matrix_log_def}. Second, the step from Eq. \eqref{eq:result_L_with_U} to \eqref{eq:result_L_without_U} is not trivial anymore, since in the nondiagonalizable case the matrix $\ln(D)$ is replaced with a nondiagonal matrix. Nevertheless, one can show that the matrix $U$ still commutes with this matrix by using Theorem 1.25 in Ref. \cite{Buch_Matrix_log} (cf. also the proof of Theorem 1.26, ibid). 

\re{From a practical viewpoint, the case of a nondiagonalizable $\GMT$ can often be neglected, since perturbing a nondiagonalizable matrix slightly causes it to become diagonalizable without significantly changing numerical results. In cases where experimental limitations like limited statistics or measurement errors prevent us from discerning such small perturbations, the nondiagonalizable case is of little practical relevance. However, it is conceivable that some constraints on the transition rates are known through other means, e.g., it might be known that some eigenvalue is degenerate. In such cases, the set of nondiagonalizable matrices is not vanishingly small and cannot be neglected.}

We return to the diagonalizable case, for which a few remarks will clarify the next steps. First, we note that condition \eqref{condition:r_smaller_pi_over_dt}, which can be rearranged into $\Delta t<\pi/\escmax(L_0)$, provides a tight inequality. More precisely, for an arbitrary but fixed maximal escape rate $r$ and any $\varepsilon>0$ we can explicitly construct systems with permissible generator matrices $L_0\neq L_1$ that satisfy $\exp(\Delta t L_0)=\exp(\Delta t L_1)$ and $\escmax(L_0) = \escmax(L_1) = r$ for the interval length $\Delta t=\pi/r+\varepsilon$ between observations. One such example is shown in Appendix \ref{sec:example_bound_sharp}. 

Second and more important for the purpose of inference, we cannot directly check for condition \eqref{condition:r_smaller_pi_over_dt} assuming that only the propagators $\GMT$ can be observed. Thus, it is in principle possible that the generator obtained through \re{Eq. \eqref{eq:L_0_Z_lnD_Z_invers}} satisfies condition \eqref{condition:r_smaller_pi_over_dt} but is nevertheless not unique in the sense that there might be another permissible generator matrix $L_1$ with $\exp(\Delta t L_1) = \GMT$. There is no contradiction to the previously proven statement if other eligible generators violate condition \eqref{condition:r_smaller_pi_over_dt}. For this reason, we will replace inequality \eqref{condition:r_smaller_pi_over_dt} with a criterion that can be checked using the propagator only. 

\subsection{Operationally accessible sufficient criteria for uniqueness}

In a setting with access only to stroboscopic measurements the condition \eqref{condition:r_smaller_pi_over_dt} cannot be directly verified from the observations alone. As remarked above, the possibility of permissible generators that satisfy Eq. \eqref{eq:exp_dt_L_0_GMT} but not condition \eqref{condition:r_smaller_pi_over_dt} necessitates us to replace this condition with one that is operationally accessible. In this section we present two approaches to achieve this goal. In the first part of this section, we modify condition \eqref{condition:r_smaller_pi_over_dt} so that it only contains measurable quantities. In the subsequent Section \ref{sec:real_eigenvalues} we demonstrate that for a large class of propagator matrices, namely those containing only pairwise distinct real eigenvalues, the underlying generator is always unique.

\subsubsection{Bound on the maximal escape rate.} 
\label{sec:upper_bound_r_max}

\re{One way to ensure that the underlying generator matrix is unique is to derive upper bounds $\ubescmax \geq \escmax$ on the maximal escape rate $\escmax$ from a given propagator matrix. If we can verify the operationally accessible criterion $\ubescmax < \pi/\Delta t$ for a given spacing $\Delta t$, we can conclude that criterion \eqref{condition:r_smaller_pi_over_dt} is satisfied.}
The goal of this section is to establish such upper bounds $\ubescmax$ for the maximal escape rate $\escmax$ as a proof of principle. We also discuss potential modifications and improvements to this result.

A first, very crude estimate can be made based on the fact that the sum over all escape rates is indeed operationally accessible and given by
\begin{equation}
\label{eq:trace_log_det_G}
    \sum_i \esc{i}=-\text{Tr}(L_0)=-\frac{\ln(\text{det}(\GM))}{\Delta t},
\end{equation}
where we used the matrix identity $\text{det}(\exp(A))=\exp(\text{Tr}(A))$ that holds for any \re{square} matrices $A$. Since $\escmax \leq \sum_i r_i$, Eq. \eqref{eq:trace_log_det_G} constitutes an elementary bound on the maximal escape rate, which can be improved if we find lower bounds on the individual escape rates $r_i$.

One way to achieve an improvement is to make use of the equivalence between path weights and master equations, which is a common tool in stochastic thermodynamics (cf. e.g. Ref. \cite{seif12}). The propagator $\G{i}{i}=p(i,\Delta t|i,0)$ can be expressed as sum over the path weights of all trajectories that end in state $i$ at time $t=\Delta t$ conditioned on their start in state $i$ at time $t=0$. Since all path weights are nonnegative, this sum can be bounded from below by the path weight of the constant trajectory that remains in state $i$ for the entire time, which is given by $\exp(-\Delta t\, \esc{i})$. Thus, we obtain the inequality
\begin{equation}
\label{ineq:lower_bound_G_ii}
\G{i}{i}\geq \exp(-\Delta t\, \esc{i}),
\end{equation}
which can be rearranged into the desired lower bound
\begin{equation}
\label{ineq:lower_bound_esc_i}
    \esc{i}\geq -\frac{\ln(\G{i}{i})}{\Delta t}\equiv \lbesc{i}
.\end{equation}
A comparison between $\sum \lbesc{i}$ and $\sum_i \esc{i}$ allows us to assess the tightness of this lower bound. Let us define the total difference between the actual escape rates and their corresponding bounds as
\begin{equation}
    d\equiv \sum_i(\esc{i}-\lbesc{i}) = -\frac{\ln(\text{det}(\GM))}{\Delta t} -\sum_i \lbesc{i}
,\end{equation}
where the second equality follows from Eq. \eqref{eq:trace_log_det_G}. The quantity $d$ can be determined from the propagator $\GM$ and the interval length $\Delta t$. Thus, we obtain upper bounds
\begin{equation}
    \esc{i}\leq \lbesc{i}+d\equiv \ubesc{i},
\end{equation}
on each individual escape rate $r_i$, which imply the operationally accessible upper bound
\begin{equation}
    \escmax\leq \max_i(\ubesc{i})\equiv \ubescmax
\end{equation}
on the maximal escape rate. Thus, for generators that satisfy \eqref{eq:exp_dt_L_0_GMT} the inequality 
\begin{equation}
\label{ineq:operationally_accessible_criterion}
  \ubescmax < \frac{\pi}{\Delta t}
\end{equation}
is a sufficient criterion for condition \eqref{condition:r_smaller_pi_over_dt}. Making use of the result from Section \ref{sec:small_dt}, we can then conclude that there is only one generator that leads to the observed $\GM$, which is given by Eq. \eqref{eq:L_0_Z_lnD_Z_invers}, provided that condition \eqref{ineq:operationally_accessible_criterion} is satisfied.

We can find a bound that is slightly superior to the result \eqref{ineq:operationally_accessible_criterion} when recalling the reasoning in Section \ref{sec:small_dt} (cf. also Figure \ref{fig:proof_different_scenarios_gerschgorin_circles}). The generator obtained from Eq. \eqref{eq:L_0_Z_lnD_Z_invers} is guaranteed to be unique if we can establish that $|\Im \mu_i|<\pi/\Delta t$ for every eigenvalue $\mu_i$ of a permissible generator matrix. While the imaginary part of the eigenvalues is not operationally accessible, their real part $\Re \mu_i$ is well-defined if the propagator is known (cf. Eq. \eqref{eq:matrix_log_def}). For this reason, the following bound
\begin{equation}
\label{ineq:operationally_accessible_criterion_2}
    \frac{\pi}{\Delta t} > \max_i \sqrt{(\ubescmax)^2 - (\ubescmax - |\Re \mu_i|)^2}
\end{equation}
still contains accessible quantities only while offering a slight improvement over Eq. \eqref{ineq:operationally_accessible_criterion}. Thus, if we can verify either inequality \eqref{ineq:operationally_accessible_criterion_2} or the simpler bound \eqref{ineq:operationally_accessible_criterion}, the generator $L_0$ is uniquely determined by Eq. \eqref{eq:L_0_Z_lnD_Z_invers}. The results \eqref{ineq:operationally_accessible_criterion_2} and \eqref{ineq:operationally_accessible_criterion} have been first derived by Cuthbert in Ref. \cite{cuth73} in Theorem 2 and its corollary, respectively. We refer to this work for a detailed proof of Eq. \eqref{ineq:operationally_accessible_criterion_2}. The same reference also points out that ``[Eq. \eqref{ineq:operationally_accessible_criterion_2}] is not the only approach possible'' to obtain even stronger bounds. In the following, we briefly sketch one such approach that refines the reasoning via path weights that led to Eq. \eqref{ineq:lower_bound_G_ii}. \re{To our knowledge, this result has not been reported explicitly in the mathematical literature before.}

To improve on the previously established upper bound $\ubescmax$, we note that we can use the known matrix $\GM$ to bound the unknown rates $\kij$ from below and above as shown in Appendix \ref{sec:Derivation_bounds_transition_rates}. The derivation is conceptually similar to the one for the bounds on the escape rates. With a lower bound on the transition rates, we can improve the bound $\ubescmax$ by taking into account more paths than the constant trajectories that were used to establish Eq. \eqref{ineq:lower_bound_G_ii}.

For instance, we may include trajectories $\gamma$ with two jumps that start in state $i$, jump to a state $j\neq i$ at time $t_1$, then jump back into state $i$ at time $t_2$ and remain there until the final time $\Delta t$. The probability $P_j$ that one of these trajectories occurs is given by integrating the path weight of $\gamma$ over the jump times $t_1$ and $t_2$. The unknown transition rates $\kij$ and $\kji$ and escape rates $\esc{i}$ and $\esc{j}$ that enter $P_j$ are then estimated using the corresponding lower and upper bounds, respectively. This yields an operationally accessible lower bound $P_j^{\text{lb}}$ for the probability $P_j$, which implies
\begin{equation}
    \G{i}{i}\geq \exp(-\Delta t \esc{i})+\sum_{j\neq i}P_j^{\text{lb}}
\end{equation}
as a refinement of Eq. \eqref{ineq:lower_bound_G_ii}. We can then follow the same steps as above to obtain a tighter version of the bound \eqref{ineq:operationally_accessible_criterion}.

\subsubsection{Real eigenvalues.}
\label{sec:real_eigenvalues}

If the matrix $\GMT=\exp(\Delta t L_0)$ has only real, pairwise distinct eigenvalues, the matrix $L_0$ can be determined exactly from $\GM$ and is given by Eq. \eqref{eq:L_0_Z_lnD_Z_invers}. This result is independent of the value of $\Delta t$. Further details on this result and other relationships between the spectral structure of $\GMT=\exp(\Delta t L_0)$ and the problem of identifying the associated generator are discussed in more detail in Ref. \cite{sing76}.

The proof is similar to the one in Section \ref{sec:small_dt}. The difference is that we can set $J=(0,...,0)$ immediately in Eq. \eqref{eq:matrix_log_def}. The reason is that, since all eigenvalues of $\GM$ are positive, the eigenvalues of $L_0$ are real and distinct as well. For any $J\neq (0,...,0)$, however, the resulting $L_J$ would contain complex eigenvalues, but since the real part of every eigenvalue is unique, eigenvalues of $L_J$ cannot occur in complex conjugate pairs. Thus, $L_J$ cannot be a real matrix for $J\neq (0,...,0)$, so that  we can proceed as in Section \ref{sec:small_dt} from Eq. \eqref{eq:result_L_with_U}.

The following example shows that the condition of distinct eigenvalues is necessary. An illustration of this example is given in Figure \ref{fig:example_two_different_generators}. We consider the generator matrix 
\begin{equation}L_0=
    \begin{bmatrix}
        -2k&k&k\\
        k&-2k&k\\
        k&k&-2k
    \end{bmatrix},
\end{equation}
which has the eigenvalues $\mu_0=0$ and $\mu_1=\mu_2=-3k$. The matrix exponential $\GMT=\exp(\Delta t L_0)$ has the eigenvalues $\lambda_0=1$ and $\lambda_1=\lambda_2=e^{-3 k\Delta t}$, i.e. real eigenvalues of which two are identical. We write the matrix in the form $L_0=W\text{diag}(0,-3k,-3k)W^{-1}$ and consider the matrix
\begin{equation}
L_1= W\text{diag}\left(0,-3k+\frac{2\pi i}{\Delta t},-3k-\frac{2\pi i}{\Delta t}\right)W^{-1}.
\end{equation}
This matrix satisfies the equation
\begin{equation}
    e^{\Delta t L_1}=e^{\Delta t L_0}=\GMT.
\end{equation}
Explicit calculation yields
\begin{equation}
    L_1=\begin{bmatrix}
        -2k&k+\frac{c}{\Delta t}      &   k-\frac{c}{\Delta t}    \\
          k-\frac{c}{\Delta t}  &-2k   &  k+\frac{c}{\Delta t}      \\
        k+\frac{c}{\Delta t}    &  k-\frac{c}{\Delta t}     &-2k 
    \end{bmatrix}\neq L_0
\end{equation}
with $c\equiv 2\pi/\sqrt{3}$.

If $\Delta t>c/k$, $L_1$ is a permissible generator matrix. Thus, for each sufficiently large interval $\Delta t>c/k$ stroboscopic observations of the generators $L_0$ and $L_1$ cannot be distinguished. This example also illustrates that in the general case stroboscopic observations cannot distinguish equilibrium and nonequilibrium. We note that this example does not stand in contradiction to the results of Sections \ref{sec:small_dt} and \ref{sec:upper_bound_r_max}, since $\Delta t>c/k>\pi/(2k)=\pi/\escmax$ is necessary for $L_1$ to qualify as a permissible generator matrix.

For systems with two states, the generator matrix has one nonzero eigenvalue, which must be real, thus the result for real eigenvalues also shows that in the simple case of two-state systems, the generator matrix can always be determined from $\GM$ independently of $\Delta t$. More generally, it has been shown that two different equilibrium systems with the same $\GM$ cannot exist \cite{Equilibrium_embedding_problem}.

\subsection{More than one possible generator}
\label{sec:arbitrary_dt_imag_eigenvalues}

In the most general case, multiple permissible generator matrices with the same matrix exponential $\exp(\Delta t L_J)=\GMT$ can exist. As the proof in Section \ref{sec:small_dt} shows, every possible generator matrix has the form of Eq. \eqref{eq:matrix_log_def}. However, the converse is not true: Not every matrix of this form qualifies as a permissible generator matrix, e.g., if it has negative off-diagonal entries. 
In this section, we discuss the generic case that the eigenvalues of $\GM$ are not degenerate, which allows for more than one but finitely many possible underlying generators in the general case of arbitrary $\Delta t$.

The results of Section \ref{sec:upper_bound_r_max} provide an upper bound on the maximal escape rate of all possible underlying generator matrices, which in turn limits the maximal imaginary part of all eigenvalues through Gerschgorin's Circle Theorem. Following the notation and reasoning of Section \ref{sec:small_dt}, the $j_i$ are bounded by (cf. Eqs. \eqref{eq:sec2:mu_i} and \eqref{inequ:IM_mu_geq_pi_over_dt}) 
\begin{equation}
\label{inequ:bound_j_i_arbitrary_dt}
    |j_i|\leq \frac{1}{2}\left(\frac{\Delta t}{\pi}\ubescmax+1\right)
.\end{equation}
\re{Since we assume that the eigenvalues $\lambda_i$ of $\GM$ are non-degenerate and the matrix $U$ in Eq. \eqref{eq:matrix_log_def} has to commute with the matrix $D=\text{diag}(\lambda_1,...,\lambda_N)$, $U$ has to be diagonal as well. Thus, $U$ also commutes with the diagonal matrix in brackets in Eq. \eqref{eq:matrix_log_def} and can therefore be ignored.}
Thus, all candidates for permissible generator matrices that solve $\exp(\Delta t L_J)=\GMT=ZDZ^{-1}$ are given by
\begin{equation}
    L_J=Z\left(\frac{1}{\Delta t} \ln(D)+\frac{2\pi i}{\Delta t}\text{diag}(j_1,...,j_N)\right)Z^{-1}
    \label{eq:ch3:potentail_generators}
\end{equation}
with $j_i\in\mathbb{Z}$ satisfying inequality \eqref{inequ:bound_j_i_arbitrary_dt}. In particular, the number of potential permissible generator matrices is finite. It has been shown \cite{cuth73} that the number of potential generator matrices either remains one for all times $\Delta t$ or tends to infinity as $\Delta t \to \infty$, at least as long as the case of degenerate eigenvalues in $\GMT$ is excluded. In the peculiar case that the geometric multiplicity of an eigenvalue exceeds one, it is possible to have uncountably many candidate generators due to the additional degrees of freedom in the matrix $U$ in Eq. \eqref{eq:matrix_log_def}.

In preparation for Section \ref{sec:4}, we now discuss in more detail how to obtain bounds on thermodynamic quantities like entropy production and cycle affinities in the case of finitely many possible underlying generators. First, we have to check which of the matrices $L_J$ of the form of Eq. \eqref{eq:ch3:potentail_generators} actually are permissible generator matrices and subsequently remove all candidates that are not. In a second step, we can compute all possible values of the quantity of interest explicitly and identify upper and lower bounds by taking the largest or smallest realized value within this set, respectively.

Regarding the first step, we first have to confirm that the matrix in question is real, i.e., does not contain any complex-valued entries. Under the assumption that the eigenvalues of $\GM$ are not degenerate, the matrix $L_J$ is real if and only if the eigenvalues of $L_J$ are real or occur in complex conjugate pairs, as we prove in Appendix \ref{sec:proof_arbitrary_dt_imag_eigenvalues}. It is also simple to check whether a candidate $L_J$ is column-stochastic \re{(cf. Eq. \eqref{eq:def_column_stochastic})}. The matrix $\ln(D)$ has exactly one zero eigenvalue, we assume $\ln(D)_{11}=0$ here without loss of generality. As long as we choose $j_1=0$, the resulting matrix $L_J$ is column-stochastic, as we prove in Appendix \ref{sec:proof_arbitrary_dt_imag_eigenvalues}. The final property that a candidate $L_J$ has to satisfy is that all off-diagonal entries are nonnegative, which cannot be checked directly from the spectrum of $L_J$. Thus, this property must be confirmed individually for all column-stochastic real matrices $L_J$ of the form \eqref{eq:ch3:potentail_generators}.

After removing all matrices $L_J$ that are not permissible generator matrices, we arrive at a list of finitely many possible underlying generators. We note that under stroboscopic observations, i.e., given only the propagator matrix $\GMT$, it is inherently impossible to pinpoint the true underlying generator, because all matrices produce the same statistics under stroboscopic observations if the spacing of observations $\Delta t$ is fixed. Nevertheless, it is straightforward to turn the list of potential generators into bounds on quantities that can be expressed in terms  of the transition rates and steady-state probabilities. For the purposes of thermodynamic inference, the mean entropy production rate and cycle affinities are two notable examples. By calculating the desired quantity for each potential generator, we obtain a list of values whose maximum and minimum trivially provide upper and lower bounds on the true value of the quantity in question.

\section{Comparison to complementary results in thermodynamic inference} \label{sec:4}

There are two qualitatively distinct ways to use the results of the previous section for thermodynamic inference. First, for generators with real, distinct eigenvalues or for time intervals $\Delta t$ smaller than a certain threshold, we can exactly determine the generator and hence also the mean entropy production rate and all cycle affinities.
Second, apart from the case where the eigenvalues of $\GM$ are degenerate, we obtain lower bounds on these thermodynamic quantities even above this threshold by calculating their exact values for each of the finitely many candidate generator matrices. 
In this section, we quantitatively compare these methods to extant results of thermodynamic inference in terms of their timescales and their ability to provide strong bounds on the underlying thermodynamic quantities. \re{We consider two such quantities in particular, namely entropy production and cycle affinities, but emphasize that the same approach remains applicable to infer other quantities as well.}

\subsection{Bound on entropy production}

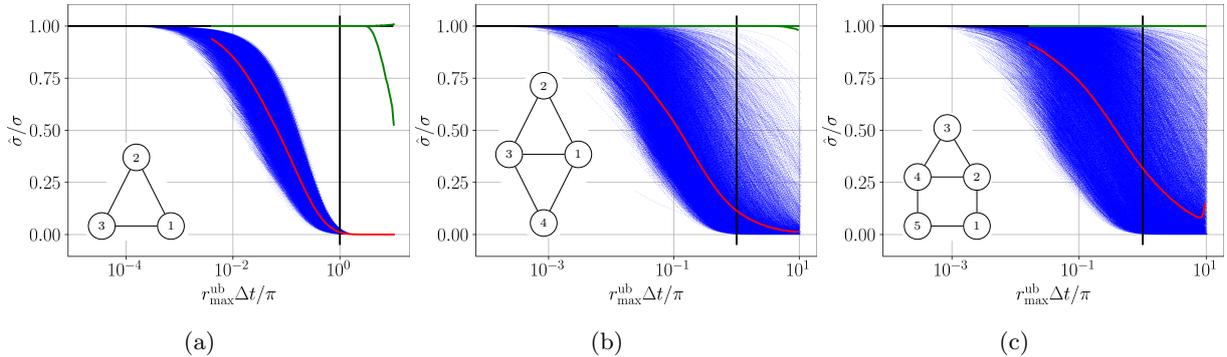
\begin{figure}[t]
\begin{minipage}{0.33\textwidth}
    \centering
    \begin{tikzpicture}
    \node at (0.95,0.95) {\includegraphics[width=\linewidth]{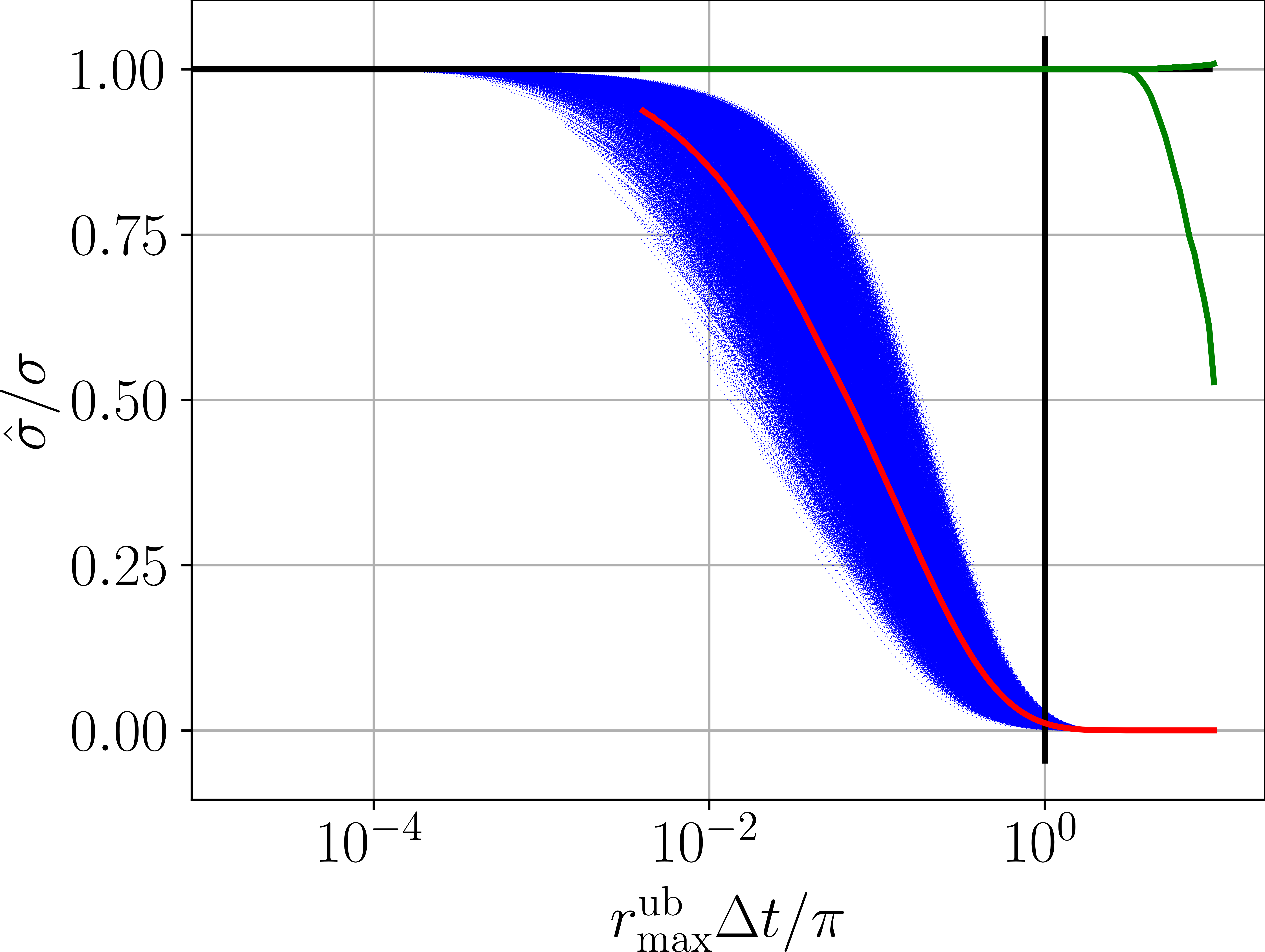}};
    \scalebox{0.65}{
        \fill[fill=white](-1.05,-0.3)--(1.1,-0.3)--(0.3,1.7)--(-0.3,1.7);

        \node[draw,circle] (1) at (0.7,0) {\scriptsize$1$};    
        \node[draw,circle] (2) at (0,1.4) {\scriptsize$2$};    
        \node[draw,circle] (3) at (-0.7,0) {\scriptsize$3$};    

        \draw (1)--(2);
        \draw (2)--(3);
        \draw (1)--(3);
    }
    \end{tikzpicture}
\subcaption{}
    \label{fig:three_state_nw_without_optimal_bound}
    \end{minipage}
    \begin{minipage}{0.33\textwidth}
    \centering
    \begin{tikzpicture}
    \node at (0.95,0) {\includegraphics[width=\linewidth]{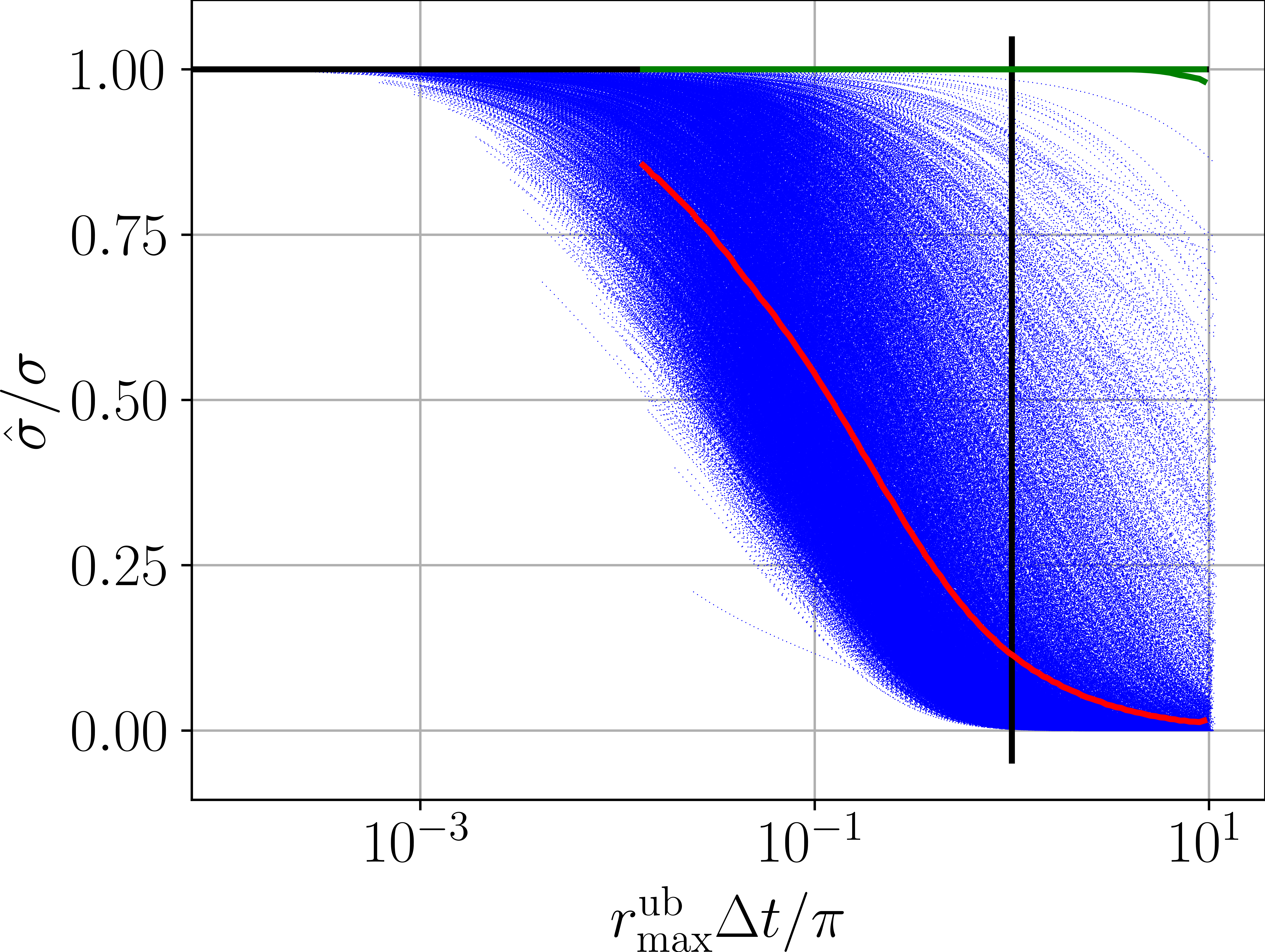}};
    \scalebox{0.65}{
    \fill[fill=white](1.0,-0.3)--(1.0,0.3)--(0.3,1.7)--(-0.3,1.7)--(-1.0,0.3)--(-1.0,-0.3)--(-0.3,-1.7)--(0.3,-1.7);

    \node[circle,draw](1) at(0.7,0) {\scriptsize$1$};    
    \node[circle,draw](2) at(0,1.4) {\scriptsize$2$};    
    \node[circle,draw](3) at(-0.7,0) {\scriptsize$3$};    
    \node[circle,draw](4) at(0,-1.4) {\scriptsize$4$};

    \draw (1)--(2);
    \draw (2)--(3);
    \draw (3)--(4);
    \draw (4)--(1);
    \draw (1)--(3);}
    \end{tikzpicture}
    \subcaption{}
    \label{fig:diamond_nw_with_optimal_bound}
\end{minipage}
\begin{minipage}{0.33\textwidth}
    \centering

    \begin{tikzpicture}
    \node at (1.0,0.95) {\includegraphics[width=\linewidth]{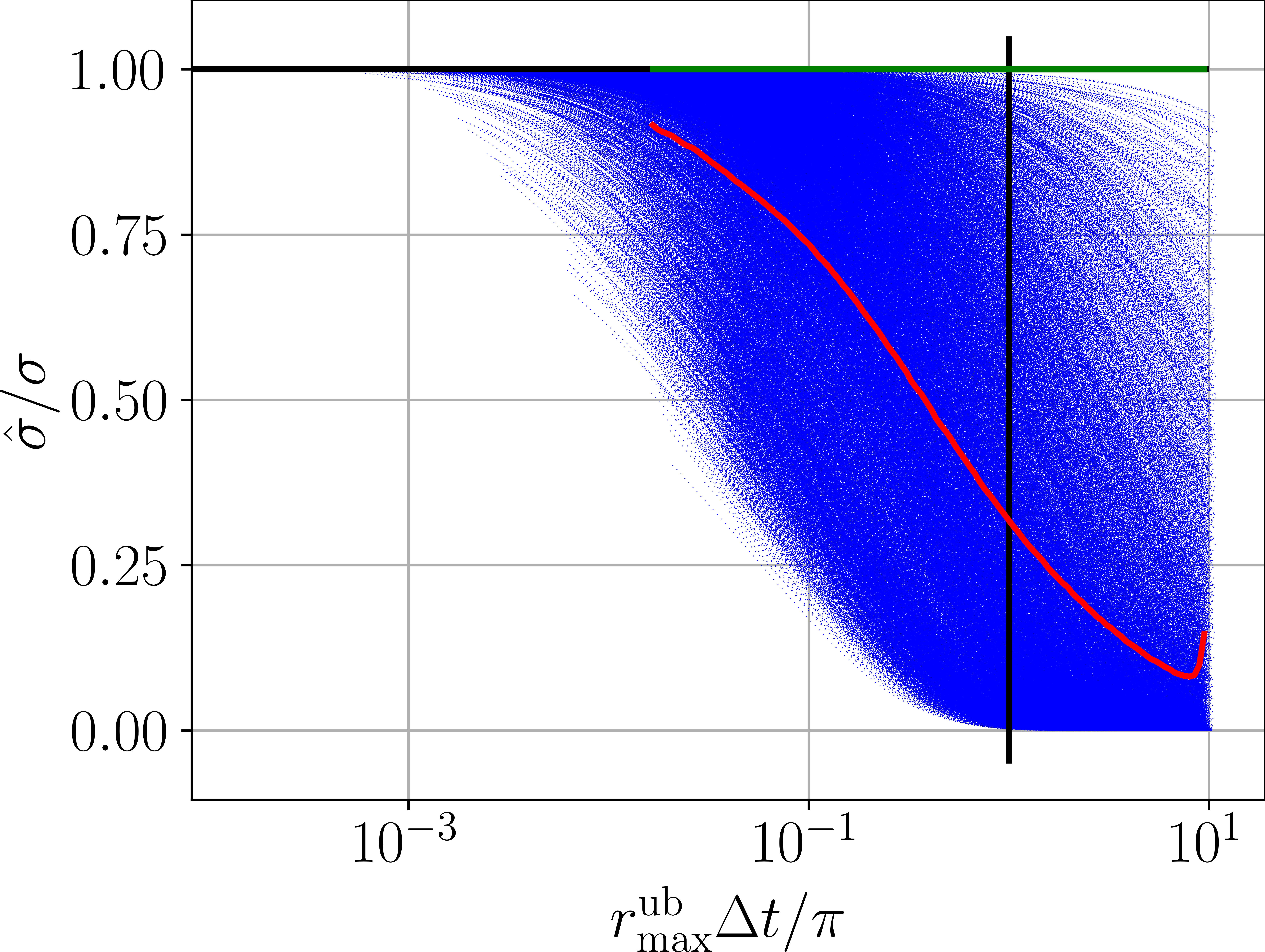}};
    \scalebox{0.65}{

    \fill[fill=white](-0.9,-0.3)--(0.9,-0.3)--(0.9,1.3)--(0.3,2.3)--(-0.3,2.3)--(-0.9,1.3);
    
    \node[circle,draw](1) at(-0.6,0) {\scriptsize$5$};
    \node[circle,draw](2) at(0.6,0) {\scriptsize$1$};
    \node[circle,draw](3) at(0.6,1) {\scriptsize$2$};    
    \node[circle,draw](4) at(0,2.0) {\scriptsize$3$};    
    \node[circle,draw](5) at(-0.6,1) {\scriptsize$4$};    

    \draw (1)--(2);
    \draw (1)--(5);
    \draw (2)--(3);
    \draw (3)--(4);
    \draw (3)--(5);
    \draw (4)--(5);
    }
    \end{tikzpicture}

    \subcaption{}
    \label{fig:house_nw_without_optimal_bound}
\end{minipage}

    \caption{Numerical illustration of entropy production estimation by reconstructing the generator and comparison to estimation based on the Kullback-Leibler divergence. The quality factors $\hat{\sigma}/\sigma$ are calculated for different transition rate configurations and interval lengths $\Delta t$ for three different network topologies, which are shown in the insets. We consider $2\cdot10^5$ different rate configurations for the three-state network in Figure (a), whereas $10^5$ configurations are used for the networks of Figure (b) and (c). Transition rates are parametrized as $\exp(-x)$, where $x$ is uniformly distributed over the interval $[-5,5]$. Since the generator and the mean entropy production rate can always be computed exactly from $\GM$ for real pairwise distinct eigenvalues, only systems with complex eigenvalues are considered in the plots. For our parametrization of transition rates, the percentage of systems with real, distinct eigenvalues is $94\%$ for the unicyclic network in (a), $89\%$ for the four-state network in (b), and $88\%$ for the five-state network in (c). For each rate configuration with at least one pair of complex eigenvalues, the quality factor $\hat{\sigma}/\sigma$ is plotted as a function of $\Delta t$, which is logarithmically spaced between $10^{-3}$ and the first value of $\Delta t$ for which $\text{det}(\GM)<10^{-15}$, with $50$ points between $10^{n}$ and $10^{n+1}$. We note that the $\Delta t$-axis is individually scaled by $\ubescmax/\pi$ for each configuration of transition rates, so that the vertical black line at $\ubescmax \Delta t/\pi = 1$ marks the value of $\Delta t$ until which we can operationally confirm that the entropy production rate is recovered exactly. Above this threshold, the green curves show the average value of the lower and upper bounds on the entropy production rate as obtained when computing all possible permissible generator matrices (cf. Section \ref{sec:arbitrary_dt_imag_eigenvalues}). As a comparison, the mean of the quality factors of the entropy estimator \eqref{eq:ch4:sigmaest} is computed in $200$ log-spaced bins between the minimum and maximum value on the horizontal axis and is shown as a red curve connecting all bins that contain at least $4\cdot10^3$ points.}
    \label{fig:scatterplots_entropy}
\end{figure}

A fairly general approach to estimate entropy production utilizes that this quantity can be expressed as a Kullback-Leibler divergence (see e.g. Ref. \cite{seif19} and references therein). Following the formalism of Ref. \cite{vdm23}, the Kullback-Leibler divergence of the path weights of forward and backward trajectories can be replaced by the simpler expression
\begin{equation}
    \sigmacgr\equiv\frac{1}{\Delta t}\sum_{ij}p^{\text{s}}_i\re{\G{j}{i}}\text{ln}\left(\frac{\re{\G{j}{i}}}{\re{\G{i}{j}}}\right)\leq \sigma
    \label{eq:ch4:sigmaest}
.\end{equation}
This quantity is a lower bound on the mean entropy production rate because stroboscopic observations yield a Markov process in discrete time on the coarse-grained level or, in the more general formalism of Ref. \cite{vdm23}, yield a series of Markovian events. The estimator $\sigmacgr$ has also been studied as a lower bound on entropy production in other contexts, e.g., in Ref. \cite{cisn23} or in Ref. \cite{skin21}, which also contains a direct proof of the inequality $\sigmacgr \leq \sigma$.

To compare the method of generator reconstruction and the bounds obtained in Section \ref{sec:arbitrary_dt_imag_eigenvalues} to the estimator $\sigmacgr$, we first plot the quality factor $\hat{\sigma}/\sigma$ for different values of the interval length $\Delta t$. For randomly chosen transition rates in three different network topologies, Figure \ref{fig:scatterplots_entropy} shows the quality factor as a series of blue dots for different values of the interval length $\Delta t$. We note that the horizontal axis is scaled by $\ubescmax/\pi$ individually for each rate configuration so that different rate configurations can be compared. With this scaling, condition \eqref{ineq:operationally_accessible_criterion} is satisfied for all values on the left side of the vertical line $\Delta t \ubescmax/\pi = 1$ and ensures that reconstructing the generator exactly is possible. In this case, the full entropy production is recovered, which is visualized by the vertical black line. 
As a comparison, the red curve shows the mean value of the quality factor $\hat{\sigma}/\sigma$, which is obtained by binning configurations of transition rates and interval lengths depending on their value of $\Delta t \ubescmax/\pi$, i.e., their position on the horizontal axis.

The green curves indicate the bounds on entropy production that can be obtained from the methods described earlier in this work. For values of $\Delta t$ that satisfy $\Delta t \ubescmax / \pi < 1$, exact reconstruction of the generator yields the exact entropy production rate, i.e., the green curve is a constant with value $1$ until hitting the vertical black line. Beyond this line, the green curves depict the lower and upper bounds on entropy production, which are obtained from the method described in Section \ref{sec:arbitrary_dt_imag_eigenvalues}.

As discussed in Section \ref{sec:real_eigenvalues}, the generator can be recovered uniquely from $\GM$ for all $\Delta t$ if the generator matrix of a system only has real, pairwise distinct eigenvalues. Since in this case the mean entropy production rate can be determined exactly for all values of $\Delta t$, we exclude such systems from the plot. Instead, the fraction of systems with real, pairwise distinct eigenvalues is given for \re{each} respective network and ensemble in \re{the figure caption of} Figure \ref{fig:scatterplots_entropy}. 

For the remaining systems, we first confirm that the eigenvalues of $\GM$ are non-degenerate. If for some system and some value of $\Delta t$ the eigenvalues of $\GM$ are degenerate, i.e., if none of the conditions from Sections \ref{sec:real_eigenvalues} and \ref{sec:arbitrary_dt_imag_eigenvalues} are satisfied, the entropy estimator should be formally set to zero, but such systems did not occur in the simulation for the chosen parametrization of transition rates. After this step, we are left with generator matrices with nondegenerate eigenvalues, for which the results from Section \ref{sec:arbitrary_dt_imag_eigenvalues} can be utilized. We consider all $L_J$ that give rise to real matrices according to Section \ref{sec:arbitrary_dt_imag_eigenvalues} and that can not be ruled out by inequality \eqref{inequ:bound_j_i_arbitrary_dt}. Since we have proven that $L_J$ is column-stochastic as long as the zero eigenvalue of $L_0$ is not modified, we only have to explicitly confirm that all non-diagonal entries are non-negative and that there are no unidirectional links. We note that for our simulations the true underlying generator $L_0$ is available, thus we can compute all other candidate generators $L_J$ directly from the randomly generated $L_0$ rather than $\GM$, which reduces numerical errors. We obtain potential generators $L_J$ by adding integer multiples of $2\pi i/\Delta t$ to the eigenvalues of $L_0$ and then checking whether the resulting matrix is indeed a permissible generator matrix. The procedure described in Section \ref{sec:arbitrary_dt_imag_eigenvalues} then yields upper and lower bounds on entropy production for each individual system. The result is depicted in Figure \ref{fig:scatterplots_entropy} as a pair of green curves that illustrate how the bounds perform for different values of $\Delta t \ubescmax/\pi$. 

As a final remark, we point out that the operationally accessible value $\pi/\ubescmax$ serves as a critical threshold for the time interval $\Delta t$. This value indicates the maximal spacing between stroboscopic observations until which entropy production can be exactly recovered, provided that the data is sufficient. This value can be compared to a timescale that can be extracted from the estimator $\sigmacgr$ and roughly corresponds to the region of $\Delta t$ where this estimator decreases most steeply, we refer to Ref. \cite{cisn23} for details. Our numerical findings indicate that in many cases this timescale is smaller or of a similar magnitude than $\pi/\ubescmax$, but, as can be seen from Figure \ref{fig:scatterplots_entropy} (c), this is not always the case. We note that there is a qualitative difference between the timescale that can be extracted from $\sigmacgr$ and the threshold value $\pi/\ubescmax$. While the former indicates a transition from timescales at which $\sigmacgr$ is close to $1$ to those where this estimator approaches zero, the latter marks the transition from small timescales where $\sigma$ is fully recovered to larger timescales where a lower bound still captures a significant amount of entropy production. 

\subsection{\re{Bound on the maximal cycle affinity}}

\begin{figure}[t]
    \centering
    \includegraphics[width=0.33\linewidth]{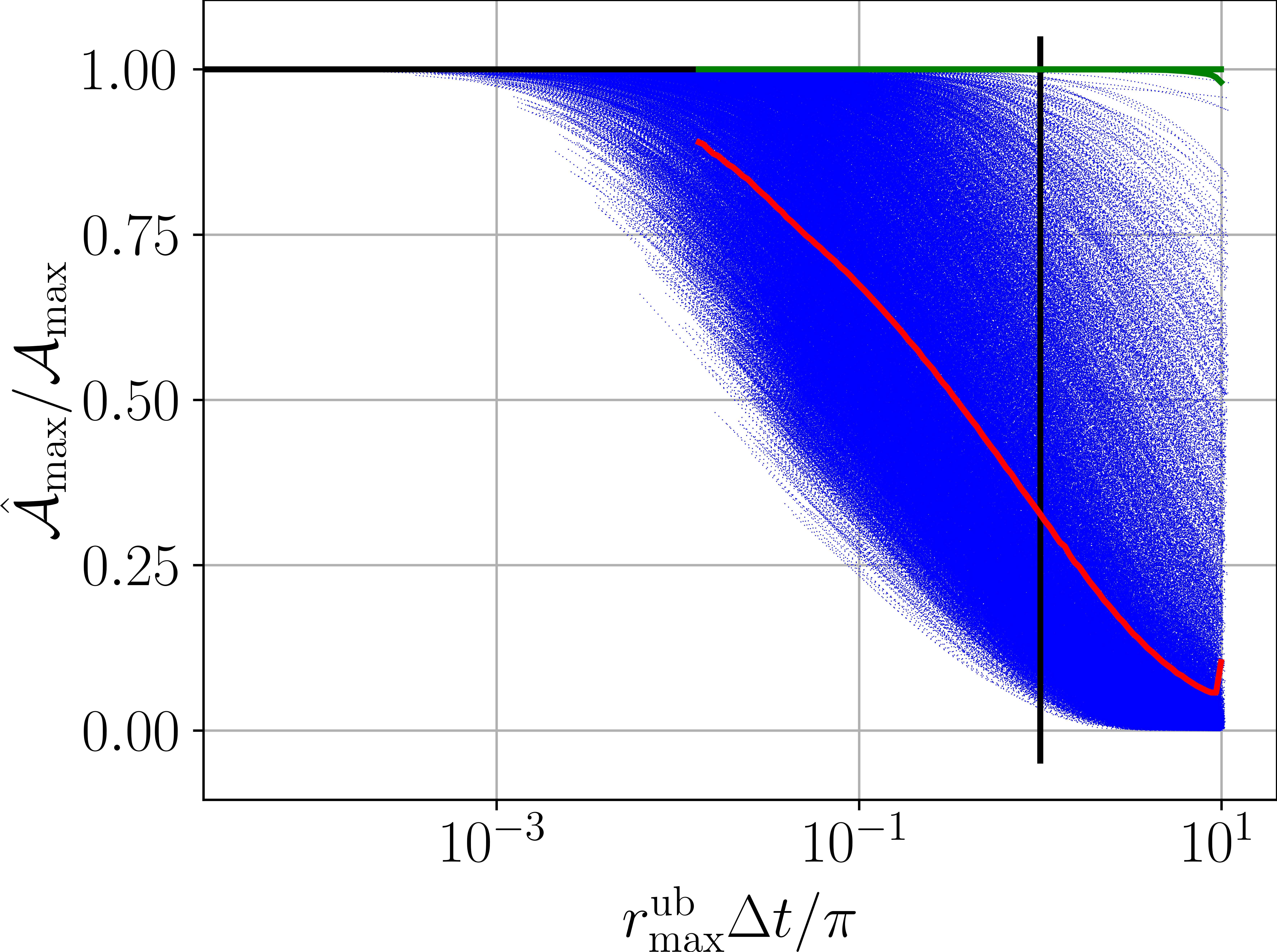}
    \caption{Numerical illustration of affinity estimation by reconstructing the generator and comparison to estimation based on \re{the conjectured} Eq. \eqref{inequ:bound_affinity}. We plot quality factors $\hat{\mathcal{A}}_{\text{max}}/\mathcal{A}_{\text{max}}$ as a function of the interval length $\Delta t$ for different rate configurations in the four-state network of Figure \ref{fig:diamond_nw_with_optimal_bound}. We consider $10^5$ different configurations of transition rates following the same distribution as in Figure \ref{fig:scatterplots_entropy}. Again, we do not include configurations with real and pairwise distinct eigenvalues in the plot, because in this case the generator can always be reconstructed exactly. For our parametrization of transition rates, the percentage of systems with real eigenvalues is $89\%$. As in Figure \ref{fig:diamond_nw_with_optimal_bound}, the $\Delta t$-axis is individually scaled by $\ubescmax/\pi$ for each configuration of transition rates, so that the vertical black line at $\ubescmax \Delta t/\pi = 1$ marks the value of $\Delta t$ until which we can operationally confirm that the affinity is recovered exactly.  Above this threshold, the green curves show the average value of the lower and upper bounds on the highest cycle affinity as obtained when computing all possible permissible generator matrices (cf. Section \ref{sec:arbitrary_dt_imag_eigenvalues}). As a comparison, the mean of the quality factors of the affinity estimator \eqref{inequ:bound_affinity} is computed in $200$ log-spaced bins between the minimum and maximum value on the horizontal axis and is shown as a red curve connecting all bins that contain at least $4\cdot10^3$ points.}
    \label{fig:scatterplot_affinity}
\end{figure}

Recent advances in thermodynamic inference consider bounds on not only entropy production but also affinities of driving cycles for different notions of partial information like transitions \cite{vdm22}, cross-correlations \cite{ohga23}, or states \cite{lian23}. \re{In the particular scenario of stroboscopic measurements, Ref. \cite{lian23} reports a conjectured bound on the sum over all cycle affinities in the system in terms of the propagators $\G{i}{j}$, which is supported by numerical evidence for small networks of size $N \leq 4$. The bound takes the form}
\begin{equation}
\label{inequ:bound_affinity:aux1}
    \ln\left(\prod_{(ij)\in\mathcal{C}_0}\frac{\re{\G{j}{i}}}{\re{\G{i}{j}}}\right)\leq\sum_{\mathcal C}\mathcal|{A}_\mathcal{C}|
,\end{equation}
\re{where $\mathcal{C}_0$ is a cycle in the system that contains four states at most}. We note that $\mathcal{C}_0$ need not be a valid cycle in the topology of the underlying Markov network; for example, $\mathcal{C}_0 = (1,2,4, 1)$ is a valid choice for the four-state network depicted in Figure \ref{fig:diamond_nw_with_optimal_bound}. 
We investigate this bound for the four-state network of Figure \ref{fig:diamond_nw_with_optimal_bound}. \re{For this particular network topology, our own numerical evidence (cf. Figure \ref{fig:scatterplot_affinity}) suggests that the bound can be improved by a factor of $1/2$, which leads to}
\begin{equation}
\label{inequ:bound_affinity}
       \hat{\mathcal{A}}_{\text{max}}\equiv\max_{\mathcal{C}}\left| \ln\left(\prod_{(ij)\in\mathcal{C}}\frac{\re{\G{j}{i}}}{\re{\G{i}{j}}}\right)\right| \leq \re{\frac{1}{2} \sum_{\mathcal C}\mathcal|{A}_\mathcal{C}| =} \max_{\mathcal C}|\mathcal{A}_\mathcal{C}|\equiv \mathcal{A}_{\text{max}}.
\end{equation}
\re{The equality $ \sum_{\mathcal C}\mathcal|{A}_\mathcal{C}| = 2 \max_{\mathcal C}|\mathcal{A}_\mathcal{C}|$ holds true for the specific network topology of Figure \ref{fig:diamond_nw_with_optimal_bound} comprising three cycles in total, two three-state cycles and one four-state cycle with an affinity equal to the sum of the previous two.} 

We now compare this bound to the results from Sections \ref{sec:real_eigenvalues} and \ref{sec:arbitrary_dt_imag_eigenvalues}. We show the results in Figure \ref{fig:scatterplot_affinity}, which depicts the quality factor of the affinity estimator \eqref{inequ:bound_affinity} for different rate configurations and different $\Delta t$ for the four-state network of Figure \ref{fig:scatterplots_entropy} (b). As in the case of entropy estimation, we compare this estimator to our proposed method in a plot similar to Figure \ref{fig:scatterplots_entropy}. All cycle affinities are recovered exactly for values of $\Delta t$ smaller than an operationally accessible threshold, which is normalized to $1$ in the figure. Above this threshold, we obtain lower and upper bounds on the cycle affinities of each individual cycle. We emphasize that this is a qualitatively different result to the estimator \eqref{inequ:bound_affinity}, which only provides a single lower bound on the largest cycle affinity rather than individual upper and lower bounds for every cycle.

\section{Discussion and conclusion}
\label{sec:5}

\paragraph{Optimality of the proposed estimator and possible refinements.} In this work we assume a stroboscopically observed continuous-time Markov process and propose the exact reconstruction of its generator as a tool for thermodynamic inference in this scenario. We provide operationally verifiable criteria under which we can explicitly recover the underlying generator and guarantee its uniqueness. In this scenario, all rates can be determined from the observation, which in turn allows to calculate thermodynamic quantities of interest like entropy production or cycle affinities. In this case, the problem of thermodynamic inference is solved completely. We have also sketched how the sufficient operationally accessible criteria on the spacing $\Delta t$ can be improved further. In particular and as pointed out in Section \ref{sec:upper_bound_r_max}, we can in principle find more refined criteria that guarantee uniqueness of the generator from weaker assumptions.

Our method also demonstrates that even when we cannot recover the generator uniquely, we obtain a list of finitely many candidates in the generic case that the eigenvalues of $\GM$ are nondegenerate.
From the perspective of thermodynamic inference,  we obtain lower and upper bounds on the desired quantity by selecting the minimal and maximal value from the possible realizations of the generator, respectively. Thus, it is also clear that the bounds are tight, because the configuration of transition rates that saturates the bound is obtained explicitly. Moreover, this generator matrix cannot be ruled out based on the available data, even if one tries to adopt other methods of inference. In particular, stronger bounds on the maximal escape rate of generator matrices may aid us in eliminating more candidates a priori and therefore improve the speed of the algorithm described in Section \ref{sec:arbitrary_dt_imag_eigenvalues}, but cannot improve the actual thermodynamic bound.

\paragraph{Optimal entropy estimators in thermodynamic inference.} 

In light of the previous summary and the highlighted optimality of our estimator, it seems appropriate to discuss what notion of ``optimality'' is actually suitable for thermodynamic inference, e.g., when entropy production is estimated. One may, for example, intuitively expect that estimating entropy production based on the Kullback-Leibler divergence between the forward and reverse coarse-grained trajectory is already optimal, because such methods can, at least in principle, account for antisymmetry under time reversal along the entire coarse-grained trajectory. The power of this technique has been recently demonstrated in novel approaches that are able to include less obvious signatures of broken time-reversal symmetries in, e.g., waiting-time distributions or higher-order statistics \cite{vdm22, haru22, blom24}. On a quantitative level, these recent methods outperform more usual approaches like the TUR, as has been demonstrated numerically and even conjectured on a general level. 

Nevertheless and as demonstrated not only in this work but also earlier, e.g., in Ref. \cite{skin21}, even the best possible bound that can be obtained from a Kullback-Leibler divergence can in principle be outperformed by estimators that explicitly utilize that the underlying model is a Markov network. Informally speaking, such methods can exploit that dwell times in states are exponentially distributed, a fact to which the Kullback-Leibler divergence between forward and backward paths remains oblivious because dwell-time distributions affect only the symmetric part of the path weight. 

A minimal example is the qualitative inference of broken detailed balance in a three-state Markov model in which two states are lumped together. Investigating whether a particular dwell-time distribution in the lumped state can emerge under equilibrium conditions or requires driving can reveal broken detailed balance \cite{aman10} despite a null result when the Kullback-Leiber divergence is calculated in this scenario \cite{luce22}. Thus, the operationally important question of whether one can decide which type of entropy estimation performs best in a particular situation cannot have a universal answer even for model-free estimators. Instead, this question is inevitably tied to particular a-priori assumptions about the underlying model class that is used to describe the physical system. In this context, we mention the recent perspective article \cite{dieb24}, which also emphasizes the importance of being aware of the underlying assumptions when identifying entropy production and time reversal.

Turning the reasoning around, we can argue that, unlike the more specialized tools for Markov networks, estimation methods for entropy production based on the Kullback-Leiber divergence are robust in the sense that conceptually they are easily transferred to broader model classes like Langevin dynamics, different types of coarse graining or time-dependently driven systems \cite{vdm23, degu24}. In contrast, we do not expect that the method introduced here can be generalized to such settings in a straightforward way. For example, measurements at discrete times alone certainly do not suffice to reconstruct a time-dependently driven generator in the general case. Similarly, the nonlocal nature of the matrix logarithm prevents us from straightforward generalizations of this method to other types of coarse graining like state lumping. For the case of nonperiodic measurements, however, different methods to estimate the generator are known, as discussed, e.g., in the review Ref. \cite{metz07}.

\re{\paragraph{Practical aspects and scaling with network size.} The methods introduced in this work are demonstrated in simple but relevant examples as a proof of principle. From a practical viewpoint and with larger networks in mind, the scaling of the number of possible generators as a function of spacing $\Delta t$ and network size $N$ is a relevant factor if we follow the procedure of Section \ref{sec:arbitrary_dt_imag_eigenvalues} to derive thermodynamic bounds. A first crude estimate based on the bound \eqref{inequ:bound_j_i_arbitrary_dt} suggests that the number of potential generator matrices scales as $(\Delta t \ubescmax/\pi)^N$, since they are parametrized by the vector $J\equiv(j_1,...,j_{\az})\in \mathbb{Z}^{\az}$. Now taking into account that, first, $j_i = 0$ whenever the corresponding eigenvalue is real and simple and, second, that eigenvalues of the generator matrix must occur in complex conjugate pairs we actually obtain at most a scaling of the form
\begin{equation}
    \text{No. of generators } L_J \sim (\Delta t N \escmax/\pi)^{N_c/2}
,\end{equation}
where $N_c$ is the number of eigenvalues of the propagator $\GMT$ that are not real and simple. In this relation, we have additionally used $\ubescmax \leq N \escmax$ to obtain a scaling without implicit dependence on $\Delta t$ through the bound. Thus, although an exponential scaling in the number of states and a polynomial scaling for increasing spacing $\Delta t$ cannot be avoided, a crude estimate shows that depending on the spectral structure of the propagator a much less severe scaling is possible if the majority of the eigenvalues is real and simple. It will be an interesting subject of future research to assess how this method performs compared to other thermodynamic bounds that do not directly rely on spectral properties.}

\re{On a related note, it is worth investigating in what way additional knowledge, e.g., about the network topology or the physical origin of the driving can be incorporated into the proposed methods. A fairly general way to check for such constraints is to first reconstruct potential generators and then check whether the desired constraints are met. Since this brute-force method can easily become unwieldy, assessing the practicality of our methods and devising more efficient approaches remains a challenging open problem.}

\paragraph{The embedding problem and thermodynamic cost.} Although there seems little hope in directly transferring the generator reconstruction method to other model classes, the mathematical concepts and tools studied here might be useful to stochastic thermodynamics in less obvious ways. For example, it was believed that discrete-time Markov processes in which the stationary distribution ``is approached through a damped wave'' \cite{cole73} cannot be embedded, i.e., cannot be obtained as a stroboscopic observation of a continuous-time Markov chain. Although this claim is false in general (cf. Ref. \cite{sing76}), a common feature to counterexamples is that they are necessarily out of equilibrium due to the thermodynamic cost of such coherent oscillations \cite{bara17, ohga23, shir23b, vu23a}. It might be interesting to compare not only the spectral but also the thermodynamic properties of discrete-time Markov processes from the perspective of whether they can be embedded or not.

\section*{Acknowledgments}
We thank Julius Degünther and Tobias Maier for stimulating discussions. J.v.d.M. also thanks Jens Wirth for helpful technical remarks. J.v.d.M. was supported by JSPS KAKENHI (Grant No. 24H00833).

\begin{appendices}
\section{Tightness of condition \eqref{condition:r_smaller_pi_over_dt}}
\label{sec:example_bound_sharp}

In Section \ref{sec:small_dt} we introduce the criterion $\escmax(L_0)<\pi/\Delta t$ that ensures that there can be no other permissible generator matrix $L_1$ with $\escmax(L_1)<\pi/\Delta t$ and $\exp(\Delta t L_1)=\exp(\Delta t L_0)$. In this section, we demonstrate that the inequality is tight in the sense that for any $\Delta t$ we can find permissible generator matrices $L_0$ and $L_1\neq L_0$ with $\exp(\Delta t L_0)=\exp(\Delta t L_1)$ and $\escmax(L_0) = \escmax(L_1) = r$ as soon as the interval length is chosen slightly larger, e.g., if $\Delta t=\pi/r+\varepsilon$ for some $\varepsilon>0$.

We consider the cyclic matrix
\begin{equation}
    L_0=\begin{bmatrix}
        a_0&a_1&a_2&a_3\\
        a_3&a_0&a_1&a_2\\
        a_2&a_3&a_0&a_1\\
        a_1&a_2&a_3&a_0
    \end{bmatrix}=W\text{diag}(\lambda_0,\lambda_1,\lambda_2,\lambda_3)W^{-1}.
\end{equation}
Denoting $\rho^m=\exp(i\pi m/2)$, we can express the eigenvalues and transformation matrices as
\begin{align}
    \lambda_m & = \sum_{l=0}^3a_l\rho^{ml} \\
    \text{and } W_{jk} & = \rho^{(j-1)(k-1)}, \label{eq:app:W_form}
\end{align}
respectively. The transformation matrix satisfies
\begin{equation}
    W^{-1}=\frac{1}{4}\overline{W} \label{eq:app:W_inv}
\re{,}\end{equation}
\re{where $\overline{\,\cdot\,}$ denotes complex conjugation.} For real coefficients $a_i$, we have $\lambda_1=\overline{\lambda}_3$, whereas $\lambda_0$ and $\lambda_2$ are real. We fix $a_0=-r$ for some $r>0$. Consider an arbitrary $\varepsilon>0$ and $\Delta t=\pi/r+\varepsilon$.
We now look for $a_l >0$ with $\sum_{l=1}^3a_l=r$ for which
\begin{equation}
    L_1\equiv W\begin{bmatrix}
        \lambda_0&0&0&0\\
        0&\lambda_1+\frac{2\pi i}{\Delta t}&0&0\\
        0&0&\lambda_2&0\\
        0&0&0&\lambda_3-\frac{2\pi i}{\Delta t}
    \end{bmatrix}W^{-1}
\end{equation}
is another permissible generator matrix besides $L_0$. By construction $L_1$ satisfies $\exp(\Delta t L_1)=\exp(\Delta t L_0)$. Using Eqs. \eqref{eq:app:W_form} and \eqref{eq:app:W_inv} for $W$ and $W^{-1}$, we explicitly calculate
\begin{eqnarray}
    \left(W\begin{bmatrix}
        0&0&0&0\\
        0&\frac{2\pi i}{\Delta t}&0&0\\
        0&0&0&0\\
        0&0&0&-\frac{2\pi i}{\Delta t}
    \end{bmatrix}W^{-1}\right)_{jk} =-\frac{\pi}{\Delta t}\sin\left(\frac{\pi}{2}(j-k)\right)
\end{eqnarray}
to express the difference between $L_1$ and $L_0$ as
\begin{eqnarray}
    L_1 - L_0 = W\begin{bmatrix}
        0&0&0&0\\
        0&\frac{2\pi i}{\Delta t}&0&0\\
        0&0&0&0\\
        0&0&0&-\frac{2\pi i}{\Delta t}
    \end{bmatrix}W^{-1} =\begin{bmatrix}
        0&\frac{\pi}{\Delta t}&0&-\frac{\pi}{\Delta t}\\
        -\frac{\pi}{\Delta t}&0&\frac{\pi}{\Delta t}&0\\
        0&-\frac{\pi}{\Delta t}&0&\frac{\pi}{\Delta t}\\
        \frac{\pi}{\Delta t}&0&-\frac{\pi}{\Delta t}&0
    \end{bmatrix}
.\end{eqnarray}
In total, we obtain the simple explicit form
\begin{eqnarray}
    L_1 = L_0+\begin{bmatrix}
        0&\frac{\pi}{\Delta t}&0&-\frac{\pi}{\Delta t}\\
        -\frac{\pi}{\Delta t}&0&\frac{\pi}{\Delta t}&0\\
        0&-\frac{\pi}{\Delta t}&0&\frac{\pi}{\Delta t}\\
        \frac{\pi}{\Delta t}&0&-\frac{\pi}{\Delta t}&0
    \end{bmatrix} 
    =\begin{bmatrix}
        -r&a_1+\frac{\pi}{\Delta t}&a_2&a_3-\frac{\pi}{\Delta t}\\
        a_3-\frac{\pi}{\Delta t}&-r&a_1+\frac{\pi}{\Delta t}&a_2\\
        a_2&a_3-\frac{\pi}{\Delta t}&-r&a_1+\frac{\pi}{\Delta t}\\
        a_1+\frac{\pi}{\Delta t}&a_2&a_3-\frac{\pi}{\Delta t}&-r
    \end{bmatrix}
\end{eqnarray}
for the generator $L_1$. We set $a_1=a_2=\delta>0$ and $a_3=\frac{\pi}{\Delta t}+\delta$. Then all entries of $L_1$ are real and all non-diagonal entries are positive.
Inserting $\Delta t=\pi/r+\varepsilon$ leads to
\begin{eqnarray}
    \sum_{l=1}^3a_l=3\delta+\frac{\pi}{\pi/r+\varepsilon}\overset{!}{=}r\re{,}
\end{eqnarray}
\re{where the exclamation mark denotes that the last equality is to be fulfilled.}
This equation is solved by
\begin{equation}
    \delta=\frac{1}{3}\frac{r\varepsilon}{\pi/r+\varepsilon}>0.
\end{equation}
Since for this particular choice of $\delta$ the columns of $L_1$ and $L_0$ have the same sum, $L_1$ is also column-stochastic and therefore a permissible generator matrix. 

\begin{figure}
    \centering
    \begin{tikzpicture}[scale=1.75]
    \draw[black,->](-1.5,0)--(0.5,0);
    \draw[black,->](0,-0.9)--(0,1.1);
    \draw(0.5,0) node[anchor=north]{Re};
    \draw(0,1.1) node[anchor=east]{Im};
    \draw[blue](-0.6,0)circle(0.6);
    \draw[black](0.05,0.6)--node[anchor=west]{$0.6$}(0,0.6);
    \draw[black](0.05,-0.6)--node[anchor=west]{$-0.6$}(0,-0.6);
    \draw[black](-0.6,0)--node[anchor=south]{$-0.6$}(-0.6,0.05);
    \draw[red,<-,thick] plot coordinates{(-0.6003812437415088, 0.5988562687754737) (-0.6061949801408224, 0.5814150595775325) (-0.6116796592378233, 0.5649610222865301) (-0.6168624491257516, 0.5494126526227449) (-0.6217676072158671, 0.5346971783523986) (-0.6264168598681005, 0.5207494203956985) (-0.6308297241154425, 0.5075108276536723) (-0.6350237815313237, 0.494928655406029) (-0.6390149123443352, 0.48295526296699454) (-0.6428174963731982, 0.4715475108804052) (-0.6464445861414925, 0.4606662415755224) (-0.6499080565645785, 0.4502758303062646) (-0.6532187348260531, 0.4403437955218407) (-0.656386513436534, 0.430840459690398) (-0.6594204489617672, 0.42173865311469816) (-0.6623288484954343, 0.41301345451369714) (-0.6651193456155018, 0.40464196315349454) (-0.6677989672866002, 0.3966030981401995) (-0.6703741929430043, 0.3888774211709872) (-0.6728510067980774, 0.38144697960576746)};
    \draw[red,<-,thick] plot coordinates{(-0.6003812437415088, -0.5988562687754737) (-0.6061949801408224, -0.5814150595775325) (-0.6116796592378233, -0.5649610222865301) (-0.6168624491257516, -0.5494126526227449) (-0.6217676072158671, -0.5346971783523986) (-0.6264168598681005, -0.5207494203956985) (-0.6308297241154425, -0.5075108276536723) (-0.6350237815313237, -0.494928655406029) (-0.6390149123443352, -0.48295526296699454) (-0.6428174963731982, -0.4715475108804052) (-0.6464445861414925, -0.4606662415755224) (-0.6499080565645785, -0.4502758303062646) (-0.6532187348260531, -0.4403437955218407) (-0.656386513436534, -0.430840459690398) (-0.6594204489617672, -0.42173865311469816) (-0.6623288484954343, -0.41301345451369714) (-0.6651193456155018, -0.40464196315349454) (-0.6677989672866002, -0.3966030981401995) (-0.6703741929430043, -0.3888774211709872) (-0.6728510067980774, -0.38144697960576746)};
    \draw[red,<-,thick] plot coordinates{(-1.1992,0)(-1.0543,0)};
    \draw[red,-,thick](0.05,0.05)--(-0.05,-0.05);
    \draw[red,-,thick](-0.05,0.05)--(0.05,-0.05);
\end{tikzpicture}
    \caption{Largest Gerschgorin circle of $L_0$ as given by Eq. \eqref{eq:result_L_0_example_bound_sharp} and $L_1 = (L_0)^{\text{T}}$ for $r=0.6$. The circles are identical since $L_0$ and $L_1$ have the same maximal escape rate $r$. The red arrows show the change of the eigenvalues as $\varepsilon$ is decreased from $3$ to $0.01$. The largest eigenvalue $\lambda_0$ remains zero for every $\varepsilon$ and is marked as cross.}
    \label{fig:eigenvalues_example_sharp_bound}
\end{figure}
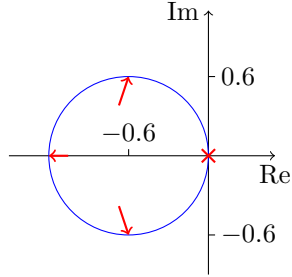

We now summarize and illustrate the essential features of this example. We can explicitly state the two generator matrices as 
\begin{equation}
\label{eq:result_L_0_example_bound_sharp}
    L_0=\begin{bmatrix}
        -r&\delta&\delta &\frac{\pi}{\Delta t}+\delta\\
        \frac{\pi}{\Delta t}+\delta&-r&\delta &\delta\\
        \delta &\frac{\pi}{\Delta t}+\delta&-r&\delta\\
        \delta &\delta &\frac{\pi}{\Delta t}+\delta&-r
    \end{bmatrix}
\end{equation}
and $L_1=(L_0)^{\text{T}}$. These matrices $L_0 \neq L_1$ are permissible generators that satisfy $\exp(\Delta t L_0)=\exp(\Delta t L_1)$ for any value $r$ of the maximal escape rate, as long as the observation interval can be written as $\Delta t=\pi/r+\varepsilon$ for any $\varepsilon>0$. Fig. \ref{fig:eigenvalues_example_sharp_bound} depicts how the eigenvalues of $L_0$ for fixed $r=0.6$ (shown as red arrows) approach the boundary of the Gerschgorin circle as $\varepsilon$ is decreased. These eigenvalues satisfy $\Im \lambda_{3,1} = \pm \frac{\pi}{\Delta t}$, so that there is a possibility to shift the eigenvalues $\lambda_1$ and $\lambda_3$ by $\pm 2\pi/\Delta t$ along the imaginary axis without moving them outside the Gerschgorin circle. 

\section{Derivation of operationally accessible bounds on individual transition rates}
\label{sec:Derivation_bounds_transition_rates}

In Section \ref{sec:upper_bound_r_max} we make use of the diagonal entries of the propagator to obtain lower bounds on the escape rate of the generator in the form $\esc{i} \geq - \ln(\G{i}{i})/\Delta t$. In this appendix we demonstrate how the off-diagonal entries of the matrix \re{$\G{j}{i}$} can be used to obtain bounds on the corresponding transition rates $k_{ij}$. 

The propagator \re{$\G{j}{i}$} can be understood as sum over the path weights of all trajectories that start in state $i$ at time $t=0$ and end in state $j$ at time $t=\Delta t$. For $i\neq j$, we obtain a lower bound on \re{$\G{j}{i}$} by considering the sum over only those trajectories that contain exactly one jump from $i$ to $j$. Using known expressions for the path weight as given in e.g. Ref. \cite{seif12}, we can formulate the corresponding inequality as
\begin{equation}
    \re{\G{j}{i}}\geq \int_0^{\Delta t}\exp(-\esc{i}\Delta t)\kij\exp(-\esc{j}(\Delta t -t)) \text{d}t
    \label{eq:app:deriv_aux1}
,\end{equation}
where the integrand is precisely the path weight of a trajectory that starts in state $i$ at time $t = 0$ and jumps to state $j$ at time $t$ to remain in this state until the final time $\Delta t$. We obtain a lower bound on the right hand side of Eq. \eqref{eq:app:deriv_aux1} by replacing the escape rates $\esc{i}$ and $\esc{j}$ by the upper bounds derived in Section \ref{sec:upper_bound_r_max}, i.e., we obtain
\begin{eqnarray}
        \re{\G{j}{i}}&\geq \kij \int_0^{\Delta t}\exp(-\ubesc{i}\Delta t)\exp(-\ubesc{j}(\Delta t -t)) \text{d}t\equiv f(\GM) \kij, 
\end{eqnarray}
where $f(\GM)>0$ is a function of the propagator and the interval length $\Delta t$. Thus, we can define 
\begin{equation}
\label{ineq:upper_bound_kij}
   \kij^{\text{ub}} \equiv \re{\G{j}{i}}/f(\GM)\geq \kij
\end{equation}
as an operationally accessible upper bound for the transition rate $\kij$. We obtain lower bounds on $\kij$ by calculating
\begin{align}
    \kij =\left(\sum_{m\neq i} k_{im}\right)-\sum_{m\neq i,j} k_{im} &=\esc{i}-\sum_{m\neq i, j} k_{im} \geq \lbesc{i}-\sum_{m\neq i, j} k_{im}^{\text{ub}}
,\end{align}
which makes use of the inequalities \eqref{ineq:lower_bound_esc_i} and \eqref{ineq:upper_bound_kij}.

\section{Proofs for Section \ref{sec:arbitrary_dt_imag_eigenvalues}}
\label{sec:proof_arbitrary_dt_imag_eigenvalues}

In this section, we use the notation of Sections \ref{sec:small_dt} and \ref{sec:arbitrary_dt_imag_eigenvalues}. In particular, there is at least one permissible generator matrix $L_0\in\mathbb{R}^{N\times N}$ that satisfies $\exp(\Delta t L_0)=\GMT$. It is now our goal to establish properties of potential generator matrices $L_J \in\mathbb{C}^{N\times N}$ parametrized by the vector of integers $J \equiv(j_1,...,j_N)$ (cf. Section \ref{sec:small_dt}).

\subsection{Proof for the characterization of real matrices}
Here, we show that $L_J$ is real if and only if the eigenvalues of $L_J$ are real or occur in complex conjugate pairs. Assuming that $L_J$ is a real matrix, it is trivial to check the desired property for the eigenvalues because the complex conjugate of an eigenvector is always an eigenvector to the complex conjugate eigenvalue. For the converse, we utilize that at least one possible underlying generator must exist, i.e., there exists at least one $J^0\equiv(j^0_1,...,j^0_N)$ such that $L_{J^0}\in\mathbb{R}^{N\times N}$. For a given $J$ we define $J^1\equiv J-J^0\equiv (j^1_1,...,j^1_N)$ and write $L_J$ as (cf. \eqref{eq:matrix_log_def})
\begin{equation}
    L_J=L_{J^0}+Z\frac{2\pi i}{\Delta t}\text{diag}(j^1_1,...,j^1_N)Z^{-1}\equiv L_{J^0}+B.
\end{equation}
We now have to show that $B$ is a real matrix.

We denote the eigenvalues of $L_{J^0}$ as $\mu_i^{J_0}$. The corresponding eigenvectors $v_i$ are also eigenvectors of $B$ to the eigenvalues $\frac{2\pi i}{\Delta t}j^1_i$, because the transformation matrix $Z$ simultaneously diagonalizes $L_{J^0}$ and $B$. We also note that the real parts of the eigenvalues of $L_{J^0}$ and $L_J$ must coincide. By assumption the eigenvalues of $L_J$ are real or occur in complex conjugate pairs and the eigenvalues of $\GM$ are non-degenerate, thus the same holds true for the eigenvalues of $L_{J^0}$. If $\mu_i^{J_0}$ is a real eigenvalue, it must be simple. In particular, there is only one eigenvalue of $L_J$ whose real part coincides with $\mu_i^{J_0}$, so we necessarily have $j^1_i=0$ and therefore also $B v_i=0$. Since $v_i$ is an eigenvector of the real matrix $L_{J^0}$ to a real, simple eigenvalue, it contains real entries only. In particular, $B v_i=0$ together with its complex conjugate implies 
\begin{equation}
    \label{eq:app:proof_aux1}
    (B -  \overline{B}) v_i=0
.\end{equation}
If $\mu_i^{J_0}\in\mathbb{C}\setminus\mathbb{R}$, $\bar{\mu}_i^{J_0}\equiv\mu_j^{J_0}$ is also an eigenvalue of $L_{J^0}$ with eigenvector $v_j=\overline{v}_i$. In this case, we must have $j^1_i=-j^1_j$ because the eigenvalues of $L_J$ also occur in complex conjugate pairs. Applying $B$ to the eigenvectors $v_i$ and $v_j$, we obtain
\begin{equation}
    B v_i=\frac{2 \pi i}{\Delta t} j^1_i v_i
\end{equation}
and
\begin{equation}
    B \overline{v}_i=B v_j=\frac{2 \pi i}{\Delta t} j^1_j v_j=-\frac{2 \pi i}{\Delta t} j^1_i \overline{v}_i
,\end{equation}
respectively. These equations imply
\begin{equation}
    \overline{B}v_i=\overline{B \overline{v}_i}=\frac{2 \pi i}{\Delta t} j^1_i v_i=Bv_i
\end{equation}
and therefore
\begin{equation}
    (B-\overline{B})v_i=0
.\end{equation}
Since $\{v_i\}_{i\in\{1,...,N\}}$ is a basis of $\mathbb{C}^N$, the previous equation together with Eq. \eqref{eq:app:proof_aux1} implies $B=\overline{B}$, i.e., $B$ is a real matrix.

\subsection{Proof for the characterization of column-stochastic matrices}
We first note that for a matrix $A=SDS^{-1}$ with corresponding diagonal matrix $D$ the $i$-th row of $S^{-1}$ is a left eigenvector of $A$ to the eigenvalue $D_{ii}$. This follows immediately from
\begin{equation}
\label{eq:left_eigenvector_column_S_invers}
    S^{-1}A=S^{-1}SDS^{-1}=DS^{-1}.
\end{equation}
A matrix is column-stochastic if and only if it has the left eigenvector $(1,1,...,1)$ to the corresponding eigenvalue zero. Since $L_0$ is a permissible generator matrix and $\exp(\Delta t L_0) =\GMT$, there has to be a $J_0\equiv (j^{0}_1,...,j^{0}_N)$ with $j^{0}_i\in \mathbb{Z}$, such that $L_{J_0}=L_0$ is a permissible generator matrix. In particular, $L_{J_0}$ is column-stochastic, i.e., $L_{J_0}$ has the left eigenvector $(1,1,...,1)$ to the eigenvalue zero. 

We now make use of the fact that the eigenspace to the eigenvalue one of $\GM$ as a propagator matrix of a connected Markov network is one-dimensional due to the Frobenius theorem. Thus, the matrix $\ln(D)$ in Eq. \eqref{eq:ch3:potentail_generators} has zero as an eigenvalue (without loss of generality as the first entry on the diagonal). Since every permissible generator matrix must have zero as a simple eigenvalue, $j^0_1=0$ follows directly.

Putting all pieces together, we can now conclude the first row of the matrix $Z^{-1}$ in Eq. \eqref{eq:ch3:potentail_generators} has to be a multiple of $(1,...,1)$. Using Eq. \eqref{eq:left_eigenvector_column_S_invers} and $j_1=0$ we see that $(1,...,1)$ has to be a left eigenvector of $L_J$ to the eigenvalue zero for any $J=(0,j_2,...,j_N)$, which implies that $L_J$ is column-stochastic.
\end{appendices}

\providecommand{\newblock}{}

\end{document}